\documentclass[twocolumn]{aastex631}

\usepackage{amsmath}
\usepackage{multirow}
\usepackage{makecell}

\usepackage{array}
\newcommand{\head}[2]{\multicolumn{1}{>{\centering\arraybackslash}p{#1}}{{#2}}}

\shorttitle{Radio-selected, optically-detected TDEs discovered in VLASS}
\graphicspath{{./}{figures/}}

\begin{document}

\title{VLASS tidal disruption events with optical flares I: the sample and a comparison to optically-selected TDEs}

\author[0000-0001-8426-5732]{Jean J. Somalwar}\email{jsomalwa@caltech.edu}
\affil{Cahill Center for Astronomy and Astrophysics, MC\,249-17 California Institute of Technology, Pasadena CA 91125, USA.}

\author[0000-0002-7252-5485]{Vikram Ravi}
\affil{Cahill Center for Astronomy and Astrophysics, MC\,249-17 California Institute of Technology, Pasadena CA 91125, USA.}

\author[0000-0001-9584-2531]{Dillon Z. Dong}
\affil{Cahill Center for Astronomy and Astrophysics, MC\,249-17 California Institute of Technology, Pasadena CA 91125, USA.}

\author[0000-0002-5698-8703]{Erica Hammerstein}
\affiliation{Department of Astronomy, University of Maryland, College Park, MD 20742, USA}

\author[0000-0002-7083-4049]{Gregg Hallinan}
\affil{Cahill Center for Astronomy and Astrophysics, MC\,249-17 California Institute of Technology, Pasadena CA 91125, USA.}

\author[0000-0002-4119-9963]{Casey Law}
\affil{Cahill Center for Astronomy and Astrophysics, MC\,249-17 California Institute of Technology, Pasadena CA 91125, USA.}

\author{Jessie Miller}
\affil{Cahill Center for Astronomy and Astrophysics, MC\,249-17 California Institute of Technology, Pasadena CA 91125, USA.}

\author{Steven T. Myers}
\affil{National Radio Astronomy Observatory, P.O. Box O, Socorro, NM 87801, USA}

\author[0000-0001-6747-8509]{Yuhan Yao}
\affil{Cahill Center for Astronomy and Astrophysics, MC\,249-17 California Institute of Technology, Pasadena CA 91125, USA.}

\author[0000-0002-5884-7867]{Richard Dekany}
\affiliation{Caltech Optical Observatories, California Institute of Technology, Pasadena, CA  91125}

\author{Matthew Graham}
\affil{Cahill Center for Astronomy and Astrophysics, MC\,249-17 California Institute of Technology, Pasadena CA 91125, USA.}

%builders
\author[0000-0001-5668-3507]{Steven L. Groom}
\affiliation{IPAC, California Institute of Technology, 1200 E. California
             Blvd, Pasadena, CA 91125, USA}
             
\author{Josiah Purdum}
\affil{Cahill Center for Astronomy and Astrophysics, MC\,249-17 California Institute of Technology, Pasadena CA 91125, USA.}

\author[0000-0002-9998-6732]{Avery Wold}
\affiliation{IPAC, California Institute of Technology, 1200 E. California
             Blvd, Pasadena, CA 91125, USA}

\begin{abstract}
In this work, we use the Jansky VLA Sky Survey (VLASS) to compile the first sample of six radio-selected tidal disruption events (TDEs) with transient optical counterparts. While we still lack the statistics to do detailed population studies of radio-selected TDEs, we use these events to suggest trends in host galaxy and optical light curve properties that may correlate with the presence of radio emission, and hence can inform optically-selected TDE radio follow-up campaigns. We find that radio-selected TDEs tend to have faint and cool optical flares, as well as host galaxies with low SMBH masses. Our radio-selected TDEs also tend to have more energetic, larger radio emitting regions than radio-detected, optically-selected TDEs. We consider possible explanations for these trends, including by invoking super-Eddington accretion and enhanced circumnuclear media. Finally, we constrain the radio-emitting TDE rate to be $\gtrsim 10$ Gpc$^{-3}$ yr$^{-1}$.
\end{abstract}

\section{Introduction} \label{sec:intro}

Extragalactic radio-synchrotron transients open a novel window onto some of the most extreme activity in the universe. These transients are typically associated with dramatic accretion events and stellar explosions, among a multitude of other possibilities. Until recently, it was impossible to obtain a uniform census of the transient radio sky due to the difficulty of performing a multi-epoch, full sky radio survey. The advent of surveys like the Caltech-NRAO Stripe 82 Survey (CNSS; \citealp{Mooley2016ApJ...818..105M}), the Jansky Very Large Array Sky Survey (VLASS; \citealp{Lacy2020}), the Australian Square Kilometre Array Pathfinder Variables and Slow Transients (ASKAP VAST; \citealp{vast}) survey have transformed our understanding by enabling the compilation of samples of radio transients that are assembled with a known selection function.

Despite this progress, many open questions remain. In particular, the relationship between radio transients and higher wavelength flares remains poorly understood, despite providing significant insight on the underlying physical processes. Tidal disruption events (TDEs) provide a quintessential example of this. TDEs occur when a star strays within the tidal radius\footnote{The tidal radius $R_T$ where the tidal forces from a SMBH overpower the internal gravity of a star is given by $R_T \approx R_* (M_{\rm BH}/M_*)^{-1/3}$ for a non-spinning black hole of mass $M_{\rm BH}$ and a star of mass $M_*$ and radius $R_*$.} of a supermassive black hole (SMBH) \citep[e.g.][]{Evans1989ApJ...346L..13E, Rees1988Natur.333..523R, Phinney1989IAUS..136..543P}. TDEs were originally theorized to produce X-ray emission, so many of the early searches for these events focused on this waveband using the ROSAT All Sky Survey (RASS; e.g., \citealp{Saxton2012A&A...541A.106S}). Soon, UV and optical searches using GALEX and SDSS began producing TDE candidates \citep[e.g.][]{Gezari2008ApJ...676..944G}. 

Throughout this effort, the question of radio emission from TDEs remained unconstrained. \cite{Giannios2011MNRAS} first suggested that a radio transient could be produced if TDEs can launch relativistic, highly collimated jets, and the first detection of a candidate jetted TDE followed shortly \citep{Bloom2011Sci}. Efforts to follow-up TDEs in the radio followed, but these tended to involve one or a few observations shortly after the higher wavelength flare and few detections resulted \citep[see][for a review]{Alexander2020SSRv..216...81A}. Recently, the detection rate has increased through long-timescale follow-up of optically-selected TDEs \citep[e.g.][]{Horesh2021,Cendes2022ApJ...938...28C,cendes_radiooptsamp}. Still, without a better understanding of the mechanisms that produce TDE radio emission and how they may be related to the multiwavelength properties of the events, it is difficult the identify the optimal candidates for follow-up, the optimal follow-up cadence, and the optimal follow-up sensitivity. 

Moreover, even if radio instruments could be used to comprehensively follow up all TDEs, the resulting radio-detected samples will be biased. Any radio-emitting TDEs discovered through follow-up of, e.g., optically-selected events, will be biased towards those TDEs that produce optical emission, and it has already been established that not all TDEs do so \citep[e.g.][]{Somalwar2021, Sazonov2021}. This renders it near impossible to compile a view of the complete landscape of radio emission through follow-up of known TDEs: if there is a correlation between optical emission and radio emission, selecting on optically-emitting TDEs will bias the expected types of radio emission.

{\addtolength{\tabcolsep}{-2pt}   \begin{deluxetable*}{cc | ccccc}
\centerwidetable
\tablewidth{10pt}
\tablecaption{Properties of our TDE sample  \label{tab:tdes}}
\tablehead{ \colhead{Name} & \colhead{AT Name} & \colhead{R.A.} & \colhead{Dec.} & \colhead{$\Delta d$ [$\arcsec$]} & \colhead{$z$} & \colhead{ $f_{\nu, \rm E2}$ [mJy]}}
\startdata
VT J0813$^{1,2,3}$ & AT 2019azh & $08^{\mathrm{h}}13^{\mathrm{m}}17.0^{\mathrm{s}}$ & $+22^\circ38{}^\prime54.0{}^{\prime\prime}$ & 0.0 & 0.022 & 1.0 \cr
VT J1008$^{4,5,6}$ & AT 2020vdq & $10^{\mathrm{h}}08^{\mathrm{m}}53.4^{\mathrm{s}}$ & $+42^\circ43{}^\prime00.2{}^{\prime\prime}$ & 0.18 & 0.045 & 1.5 \cr
VT J1356 & & $13^{\mathrm{h}}56^{\mathrm{m}}12.1^{\mathrm{s}}$ & $-26^\circ58{}^\prime50.7{}^{\prime\prime}$ & 0.54 & 0.018 & 2.5 \cr
VT J1752$^{4,6}$ & AT 2019baf & $17^{\mathrm{h}}52^{\mathrm{m}}00.1^{\mathrm{s}}$ & $+65^\circ37{}^\prime36.0{}^{\prime\prime}$ & 0.13 & 0.089 & 1.4  \cr
VT J2012 & & $20^{\mathrm{h}}12^{\mathrm{m}}29.9^{\mathrm{s}}$ & $-17^\circ05{}^\prime56.3{}^{\prime\prime}$ & 0.2 & 0.053 & 1.1  \cr
VT J2030 & & $20^{\mathrm{h}}30^{\mathrm{m}}47.3^{\mathrm{s}}$ & $+04^\circ13{}^\prime31.0{}^{\prime\prime}$ & 0.19 & 0.061 & 1.4 
\enddata
\tablecomments{Redshifts are measured from our follow-up optical spectroscopy as described in Appendix~\ref{app:optspec}. References: $^{1}$\citealp{vanVelzen2021},$^{2}$\citealp{Goodwin2022MNRAS}, $^{3}$\citealp{Sfaradi2022ApJ},$^{4}$\citealp{Yao2023},$^{5}$\citealp{ptde},$^{6}$\citealp{paperII}}
\centering
\end{deluxetable*}}

A radio-selected TDE sample is key to constraining these many unknowns, and the 3 GHz Jansky Very Large Array Sky Survey (VLASS; \citealp{Lacy2020}) is ideal for such an effort: predictions have estimated that ${\sim}100$ TDEs could be detectable using that survey \citep[][]{Anderson2020ApJ...903..116A}. The first candidates detected in this survey have already been published \citep{Ravi2021, Somalwar2021, Somalwar2023ApJ...945..142S}, and compilation of the first VLASS-selected TDE sample, regardless of multiwavelength counterpart, is underway (Somalwar et al., in prep.). 

With this VLASS TDE sample, we can begin answering some of the questions posed earlier about the range of radio emission and the emission mechanisms, as well as the relationship between the radio emission and multiwavelength emission. In this paper, we focus on the latter questions: we present the first sample of radio-selected, optically-detected TDEs.

For clarity, we have divided our analysis into three papers. In this paper (Paper I), we present our full radio-selected, optically-detected TDE sample and perform observational comparisons between radio-selected, optically-detected TDEs and optically-selected TDEs, with the aim of identifying those factors that distinguish between radio-emitting TDEs and radio-quiet TDEs. We also discuss the rate of radio-bright, optically-bright TDEs. We will briefly speculate on the physical mechanisms at play. We defer detailed discussion of both the multiwavelength properties and physical interpretations of each event to the companion papers. \cite{paperII} discusses late-time, transient, optical spectral features detected from two of the events. \cite{paperIII} discusses evidence for and implications of jet launching from three of the events.

\section{Sample Selection}

We compile our radio-selected TDE sample using data from the Very Large Array Sky Survey (VLASS; \citealp{Lacy2020}). VLASS is an ongoing effort to observe the entire sky with $\delta > -40^\circ$ at 3 GHz for three epochs with a cadence of ${\sim}2$ years. VLASS has a $1\sigma$ sensitivity of ${\sim}0.13$ mJy/beam and a spatial resolution of ${\sim}2.5\arcsec$. Each epoch is divided into two halves. The first half of epoch one, which we denote E1.1, was observed in 2017. The second half of epoch one (E1.2) was observed 2018. E2.1 was observed in 2020, and E2.2 was observed from 2021. E3.1 is ongoing (Jan 2023-present).

We identified TDE candidates using the transient catalog of Dong et al., in prep., who identified all sources that were detected at $>7\sigma$ in E2 but not significantly detected ($<3\sigma$) in E1; i.e., this catalog contains all transients that are rising between E1 and E2. Details about the transient detection algorithm are described in \cite{Somalwar2021} and Dong et al., in prep. We select TDE candidates from this catalog using the following criteria:
\begin{enumerate}
    \item the transient is within $1\arcsec$ of the position of a source in the PanSTARRS catalog \citep[][]{Chambers2016arXiv161205560C, Flewelling2020ApJS..251....7F};
    \item the associated source is a galaxy; i.e., it is inconsistent with being a star using all public catalogs, including GAIA (to remove all objects with significant parallax; \citealp{2022arXiv220800211G, 2016A&A...595A...1G}) and the PanSTARRS star-galaxy classifier \citep[][]{Beck2021MNRAS.500.1633B};
    \item the host galaxy must show no evidence for strong AGN activity. Among the criteria used to identify AGN, we consider: the position of the source on the WISE W1-W2 and W2-W3 color diagram \citep[][]{Stern2012ApJ...753...30S}; any evidence for past optical, X-ray, or radio variability/detections that could indicate AGN activity; and any public optical spectra with broad or narrow line emission that indicate strong AGN activity. We have found that this criteria rules out strong AGN, but some weak AGN remain. In this paper, we include these objects and will discuss the implications of their AGN activity primarily in \cite{paperIII};
    \item the host galaxy must have a PanSTARRS photometric redshift $z_{\rm phot}<0.25$ \citep[][]{Beck2021MNRAS.500.1633B}. When PanSTARRS photometric redshifts are unavailable, we inspect catalog data for the galaxy to identify the redshift. If no redshift information is available, we measure a redshift using an SED fit to the galaxy photometry, following the procedure discussed in Appendix~\ref{app:optspec}. We include this redshift cut to ensure that we can obtain multiwavelength follow-up to detect counterparts and classify the host galaxies with reasonable exposure times. This redshift cut will bias us against rare TDEs; e.g., we are not sensitive to most on-axis jetted TDEs, which are generally found at $z\gtrsim0.3$. Future work will consider the higher-redshift transients.
\end{enumerate}
In summary, this selection criteria will identify nuclear radio flares in nearby non- or weakly-active galaxies. We obtained optical spectroscopy for all events to measure spectroscopic redshifts, and do not consider any source with $z_{\rm spec}>0.25$. The resulting sample has ${\sim}100$ objects. We will perform a detailed analysis of the completeness of this selection in future work, when we present the full VLASS TDE sample.

In this paper, we only consider transients with associated optical flares. We obtained forced-photometry at the position of each transient from the Zwicky Transient Facility (ZTF; $gri$ bands; \citealp{Bellm2019}) and the Asteroid Terrestrial-impact Last Alert System (ATLAS; $co$ bands; \citealp{Tonry2018}) using recommended procedures. Our sample includes every transient with an optical lightcurve with at least three $5\sigma$ detections and at least one $10\sigma$ in any bands. We chose this criteria to ensure that the flare is significantly detected, and thus amenable to quantitative analysis of the flare evolution. Note that we have already ruled out events with long-term, AGN-like variability in our TDE candidate selection. While we do not do a detailed search of optical lightcurves from the All-Sky Automated Survey for Supernovae (ASASSN; \citealp{Kochanek2017PASP}) because of limited optical color information and sensitivity, our final sample includes one source that has an ASASSN flare with simultaneous MIR observations, and thus more constraints on the physical properties of the optical flare (Appendix~\ref{app:j1356}).

The resulting sample has six objects, the properties of which are summarized in Table~\ref{tab:tdes}. These sources are named using our VLASS transient naming convention: VT J081316.97+223853.99 (henceforth VT J0813), VT J100853.44+424300.22 (VT J1008), VT J135612.14-265850.71 (VT J1356), VT J175200.13+653736.04 (J1752), VT J201229.90-170556.32 (VT J2012), and VT J203057.34+041330.97 (VT J2030). We refer to the host galaxies of these objects using the coordinates prefixed with HG (host galaxy; e.g., HG J1356, HG J1008, etc.). In plots, we often label the individual transients or host galaxies without the prefixes (e.g., J1356, J1008), except when the prefixes are necessary for clarity.

\subsection{The optically-selected TDE sample}

A key aspect of this paper is our comparison of radio-selected and optically-selected TDEs; hence, here we present our chosen optically-selected comparison samples. We consider two different samples: the \cite{Yao2023} and \cite{Hammerstein2022} samples. The \cite{Hammerstein2022} sample includes all classified TDEs discovered in the first ${\sim}3$ years of ZTF. \cite{Yao2023} presents a complete sample of TDEs with peak $g$-band mag $\lesssim 19$ during the first three years of ZTF. 

We largely compare to the \cite{Yao2023} sample. However, when considering host galaxy stellar masses, colors, or star-formation rates, or Baldwin, Phillips \& Terlevich (BPT;\citealp{Baldwin1981}) classifications, we adopt the \cite{Hammerstein2022} sample, because our methods better align with theirs. 

In both cases, we restrict our comparison to events at redshifts $z<0.1$. We apply this cut to ensure that redshift evolution doesn't bias our results: the highest redshift object in our sample is at $z=0.089$, whereas the optically-selected TDEs range to much larger redshifts.

\section{Observations and Data Reduction}

We have performed an extensive multi-wavelength follow-up campaign for all of our TDE candidates. We present a relevant subset of that data here; the full dataset for each TDE candidate is presented in the corresponding companion paper. In this section, we describe our observations and data reduction procedures. 

\subsection{Radio observations} \label{sec:radioobs}

{\addtolength{\tabcolsep}{-2pt}   \begin{deluxetable*}{c | cccc}
\centerwidetable
\tablewidth{10pt}
\tablecaption{Summary of VLA follow-up \label{tab:vla_followup}}
\tablehead{ \colhead{Name} & \colhead{MJD} & \colhead{Configuration} & \colhead{Freq. range [GHz]} & \colhead{Program (PI)} }
\startdata
% VT J0813 & & & \\\hline
VT J1008 & 59612 & B$\rightarrow$BnA & $1{-}12$ & 21B-322 (G. Hallinan) \\\hline
VT J1356 & 59626 & BnA & $1{-}12$ & 21B-322 (G. Hallinan) \\\hline
VT J1752 & 59273 & A & $1{-}12$ & 20B-393 (D. Dong) \\
 & 59632 & BnA & $1{-}12$ & 21B-322 (G. Hallinan) \\\hline
VT J2012 & 59257 & A & $1{-}12$ & 20B-393 (D. Dong) \\
 & 59271 & A & $1{-}12$ & 20B-393 (D. Dong) \\\hline
VT J2030 & 58881 & C & $1{-}18$ & 19A-013 (PI: K. Alexander) \\
 & 59130 & C & $1{-}18$ & 20A-372 (PI: K. Alexander) \\
  & 59257 & C & $1{-}12$ & 20B-393 (PI: D. Dong)
\enddata
\centering
\end{deluxetable*}}

Each TDE candidate has one or more multi-frequency radio observations from the Jansky Very Large Array. We summarize the observations for all sources except J0813 in Table~\ref{tab:vla_followup}. The reduced VLA SEDs for J0813 were published in \cite{Goodwin2022MNRAS}, so we adopt the spectra tabulated in that work. We reduced all other observations using the VLA Calibration Pipeline 2022.2.0.64  and CASA version 6.4.1 \citep{casa}. We imaged the data using standard CASA recipes, and measured the flux density of each source as the flux density of the peak pixel in a $50\times 50$ pixel box centered on the source location.

\subsection{Optical transient photometry}

We retrieve forced, difference photometry from the Asteroid Terrestrial-impact Last Alert System (ATLAS; \citealp{Tonry2018}) and Zwicky Transient Facility (ZTF; \citealp[][]{Bellm2019, Graham2019PASP..131g8001G, dekany_ztf, masci_ztf}) using the recommended procedures and automated pipelines for each survey. We use public data from the ATLAS survey, and both public and partnership data from ZTF. We load the data from both surveys using the Hybrid Analytic Flux FittEr (\texttt{HAFFET}; \citealp{haffet}) code, and then bin each lightcurve with a binsize of one day. The ATLAS photometry of J0813 use different baselines for MJD $> 58895$, leading to zeropoint offsets for this MJD range. We correct for the zeropoint offset by estimating the median non-transient flux in each filter  from those data points at MJD $> 59200$ in each and subtract that median flux from the lightcurve for MJD $> 58895$. Similarly, the ZTF host reference image of J2030 was taken while the transient was active, so it overestimates the baseline flux, and we correct for this by calculating the median flux in each band for MJD$ > 58500$ and subtracting this flux from the entire lightcurve. 

We treat J1356 separately because it is the oldest transient and is not detected by ATLAS or ZTF. It is, however, detected by the All-Sky Automated Survey for Supernovae (ASASSN; \citealp{Kochanek2017PASP}). We retrieved and processed the lightcurve using the recommended ASASSN tools \citep{Jayasinghe2019MNRAS,Shappee2014ApJ}. Like for the other events, we binned the lightcurve in one day bins.

\subsection{Host photometry}

We retrieve host photometry for each TDE using the same methods as \cite{Hammerstein2022}, to enable a like-to-like comparison with the results from that work. We refer the reader to \cite{Hammerstein2022} for a detailed description of the adopted methods. In brief, we retrieve SDSS or PanSTARRS, where SDSS is unavailable, Kron magnitudes. We also retrieve GALEX NUV and FUV photometry using recommended methods.

\subsection{Optical spectroscopy} \label{sec:optspec}

HG J0813 was previously observed as part of the Sloan Digital Sky Survey (SDSS; \citealp{Abdurrouf2022ApJS}) Spectroscopic survey \citep{Strauss2002} on MJD 52943. We retrieved the spectrum of this source from the SDSS DR17 website.

We obtained optical spectra for the all our TDE candidates except VT J0813 using the Low Resolution Imaging Spectrometer (LRIS) on the Keck I telescope. In all cases, we centered the observation on the galactic nucleus using a parallactic angle. We used the 400/3400 grism, the 400/8500 grating with central wavelength 7830, and the 560 dichroic. The resulting wavelength range was ${\sim}1300{-}10000\,{\rm \AA}$ and the resolution $R{\sim}700$. We observed VT J1008 on MJD 59676 for 20 min using the $1\arcsec$ slit with the standard star Feige 34. We observed VT J1356 on MJD 59616 for 10 min using the $1\arcsec$ slit with the standard star Feige 34. We observed VT J1752 on MJD 59260 for 10 min using the $1\farcs5$ slit with the standard star Feige 34. We observed VT J2012 on MJD 59464 for 10 min using the $1\farcs5$ slit with the standard star BD+28. We observed VT J2030 on MJD 59464 for 10 min using the $1\farcs5$ slit with the standard star Feige 34. We reduced the observations using the \texttt{lpipe} code with standard settings \citep[][]{Perley2019}.

We obtained high-resolution optical spectra for a subset of the TDE candidates using the Echellette Spectrograph and Imager (ESI) on the Keck II telescope. We used the Echelle mode for all observations. We observed VT J0813 on MJD 59874 for 20 minutes using the $0\farcs5$ arcsec slit. We observed VT J1008 on MJD 59908 for 22.5 min using the $0\farcs3$ slit. We observed VT J2012 on MJD 59876 for 20 min using the $0\farcs5$ arcsec slit. We observed VT J2030 on MJD 59874 for 20 min using the $0\farcs5$ arcsec slit. The $0\farcs5$ ($0\farcs3$) slit leads to an instrumental broadening of $\sigma_{\rm inst} = 15.8 (9.5)$ km s$^{-1}$. We reduced the spectra using the \texttt{makee} software following the default, recommended procedures for ESI data reduction.

\section{Summary of detailed transient properties}

In this paper, we focus on the properties of the radio-selected, optically-bright TDE {\it population}, rather than the individual characteristics of each source. Hence, we will primarily discuss observations that are available for our full sample and for comparison optical TDE samples; namely, the optical lightcurves and spectra, host galaxy observations from public survey data, and the radio observations. Context about the individual transients is, however, useful for interpreting the results of this paper. Thus, we begin our paper with a brief review of the properties of each transient detailed in papers II/III and other sources.\\

\noindent \textbf{VT J0813} (\citealp{paperIII, vanVelzen2021, Goodwin2022MNRAS, Sfaradi2022ApJ}): VT J0813 (otherwise known as ASASSN-19dj or AT2019azh) was first discovered as an optical transient by the ASASSN survey on Feb. 22 2019 \citep{Hinkle2021MNRAS}. It was also detected by the ZTF and ATLAS surveys. The optical lightcurve is typical of TDEs. Transient Balmer and Helium lines were detected in follow-up optical spectra, leading to the classification of this source as a TDE-H+He \citep{vanVelzen2021, Hammerstein2022}. In addition to the optical flare, this source was detected as an X-ray transient with peak luminosity $L_X=10^{43}$ erg s$^{-1}$ that brightened ${\sim}7$ months after the optical peak \citep{vanVelzen2021}. The host galaxy shows a disturbed morphology, characteristic of a recent merger or interaction \citep[][]{paperIII}.

This source was first detected as a GHz radio transient by \cite{Perez2019ATel} with the e-MERLIN telescope and was then observed by numerous radio telescopes, notably by \cite{Goodwin2022MNRAS}, who obtained multi-epoch radio SEDs and argued that the radio-emitting outflow was non-relativistic and had a very optically-thin spectral index. \cite{Sfaradi2022ApJ} further observed this source with a high cadence and argued that the radio variability was correlated with the X-ray spectral variability and luminosity in a manner similar to that observed from X-ray binaries. They used this to argue for the presence of a jet subject to the accretion state changes in the accretion disk.

\noindent \textbf{VT J1008} \citep[][]{paperII, Yao2023, ptde}: VT J1008 (AT 2020vdq) was first detected as an optical transient by ZTF on Oct. 4 2020. There is no prompt follow-up for this source, so the early time radio and X-ray behavior is unknown. It was not reported as an X-ray transient in any public X-ray surveys. Late-time (${\sim}2$ years post-optical peak) Neil Gehrels Swift Observatory X-ray Telescope (Swift/XRT; \citealp{swift_xrt}) follow-up ${\sim}2$ years post-TDE did not detect any significant X-ray emission ($3\sigma$ upper limit $<10^{41.8}$ erg s$^{-1}$). In an optical spectrum ${\sim}2$ years post-optical-peak, transient Balmer lines and He\,II lines are detected with widths ${\sim}1000$ km s$^{-1}$ and luminosities ${\sim}10^{40}$ erg s$^{-1}$. These lines are commonly detected from TDEs, but they are typically much broader (${\sim}10^4$ km s$^{-1}$) and fade within a year. The origin of these lines is considered in detail by \cite{paperII}. Roughly three years after the initial optical flare, this source re-brightened in the optical. \cite{ptde} constrain the origin of the rebrightening and suggest that this event is a repeating, partial tidal disruption event. No new radio emission was detected post-rebrightening.

\noindent \textbf{VT J1356} (Appendix~\ref{app:j1356}): VT J1356 is detected as a transient in ASASSN forced photometry. It is simultaneously detected as a NIR transient in NEOWISE photometry. It is the oldest of our TDE sample, with the ASASSN detections occuring ${\sim}8$ years ago. The NIR flare is double peaked, with an initial, hot ($T\sim10^4$ K) flare followed by a $\gtrsim5$ year cool flare ($T\sim 10^3$ K), consistent with blackbody emission from dust at ${\sim}0.5$ pc that has been heated by the high energy emission produced during the transient. This source is not discussed in either of the companion papers, so we describe its properties in detail in Appendix~\ref{app:j1356}.

\noindent \textbf{VT J1752} (\citealp{paperIII}, simultaneously reported by \cite{Yao2023}): VT J1752 (AT2019baf) was first reported as a transient by ZTF on Jan. 9 2019. The optical lightcurve showed an unusual double peaked structure. There was no prompt multiwavelength follow-up for this source, so the early time radio and X-ray behavior is unknown. It was not reported as an X-ray transient in any public X-ray surveys. Late-time (${\sim}2$ years post-optical peak) Swift/XRT follow-up detected an X-ray source with $L_X = 10^{42.4}$ erg s$^{-1}$. No transient optical spectral features were detected in follow-up optical spectra. The host galaxy shows a disturbed morphology, characteristic of a recent merger or interaction. The radio SED is best-fit by a multi-component synchrotron model, suggesting the presence of multiple outflows or a structured outflow. The outflows have high luminosities and energies, suggestive of the presence of a jet.

\noindent \textbf{VT J2012} (\citealp{paperII}): VT J2012 was first detected as a transient by the ATLAS survey on 58772. It has multiwavelength properties identical to those of VT J1008. The only major differences between these two events is that no transient Helium lines are detected from VT J2012, and the transient Balmer lines are redshifted by ${\sim}700$ km s$^{-1}$. There is no blueshifted component.

\noindent \textbf{VT J2030} (\citealp{paperIII}): VT J2030 was first detected as transient by the XMM-Newton Slew Survey team, who undertook an X-ray and radio follow-up campaign. The X-ray emission from this source has an peak luminosity ${\sim}10^{43.8}$ erg s$^{-1}$, but with strong variability of a factor of a few on ${\sim}$week timescales. The radio emission also shows non-monotonic evolution. Both of these properties are consistent with the presence of a young jet.

VT J2030 was detected as an optical transient by the ZTF survey. However, the transient occurred during the start of ZTF, so it was present in the reference images and the transient appeared as a negative flux in the ZTF photometry. VT J2030 was also detected as a transient by the ATLAS survey. In this survey, it appeared like a typical TDE, with a positive rise and then decay. 

\section{Host galaxies} \label{sec:host}

{\addtolength{\tabcolsep}{-2pt}   \begin{deluxetable*}{c | cccccccccc}
\centerwidetable
\tablewidth{10pt}
\tablecaption{Host Galaxy %\todo{real errors for $sigma_*$}
\label{tab:host}}
\tablehead{ \colhead{Name} & \colhead{$z$} & \colhead{$\log\frac{M_*}{M_\odot}$} & \colhead{$\log\frac{\sigma_*}{{\rm km s}^{-1}}$} & \colhead{$\log \frac{M_{\rm BH}(M_*)}{M_\odot}$} & \colhead{$\log \frac{M_{\rm BH}(\sigma_*)}{M_\odot}$} & \colhead{BPT} & \colhead{$\log\frac{\rm SFR_{SED}}{(M_\odot\/{\rm yr})}$} & \colhead{$A_V$} & \colhead{$(u-r)_0$}}
\startdata
J0813 & 0.022 & $9.79_{0.00}^{0.00}$ & $68 \pm 2$ & $6.99^{+0.19}_{-0.21}$ & $6.44 \pm 0.29$ & Sey. & $0.27_{0.04}^{0.04}$ & $0.12^{+0.08}_{-0.10}$ & $1.77_{0.00}^{0.01}$ \\
J1008 & 0.045 & $9.16_{0.10}^{0.30}$ & $44 \pm 3$ & $4.81^{+0.40}_{-0.32}$ & $5.59 \pm 0.29$ & Q? & $0.07_{0.06}^{0.03}$ & $0.18^{+0.14}_{-0.13}$ & $1.86_{0.11}^{0.15}$ \\
J1356 & 0.018 & $8.94_{0.09}^{0.39}$ & $-$ & $5.08^{+0.28}_{-0.30}$ & $-$ & SF & $0.08_{0.08}^{0.07}$ & $0.08^{+0.05}_{-0.23}$ & $1.63_{0.24}^{0.25}$ \\
J1752 & 0.089 & $10.07_{0.07}^{0.32}$ & $-$ & $6.91^{+0.28}_{-0.31}$ & $-$ & Sey. & $2.41_{0.78}^{3.70}$ & $1.10^{+0.35}_{-0.30}$ & $1.93_{0.15}^{0.15}$ \\
J2012 & 0.053 & $9.90_{0.09}^{0.32}$ & $59 \pm 2$ & $6.55^{+0.24}_{-0.32}$ & $6.17 \pm 0.31$ & Q? & $0.00_{0.00}^{0.06}$ &  $0.32^{+0.23}_{-0.28}$ & $2.34_{0.10}^{0.08}$ \\
J2030 & 0.061 & $9.79_{0.01}^{0.26}$ & $62 \pm 5$ & $6.37^{+0.22}_{-0.23}$ & $6.25 \pm 0.31$ & SF & $0.50_{0.07}^{0.10}$ &  $0.14^{+0.09}_{-0.13}$ & $1.84_{0.07}^{0.09}$
\enddata
\tablecomments{The host galaxy properties of our radio-selected TDE sample. The methods of measuring these properties are described in Section~\ref{sec:host}.}
\centering
\end{deluxetable*}}

\begin{figure*}
    \centering
    \includegraphics[width=\textwidth]{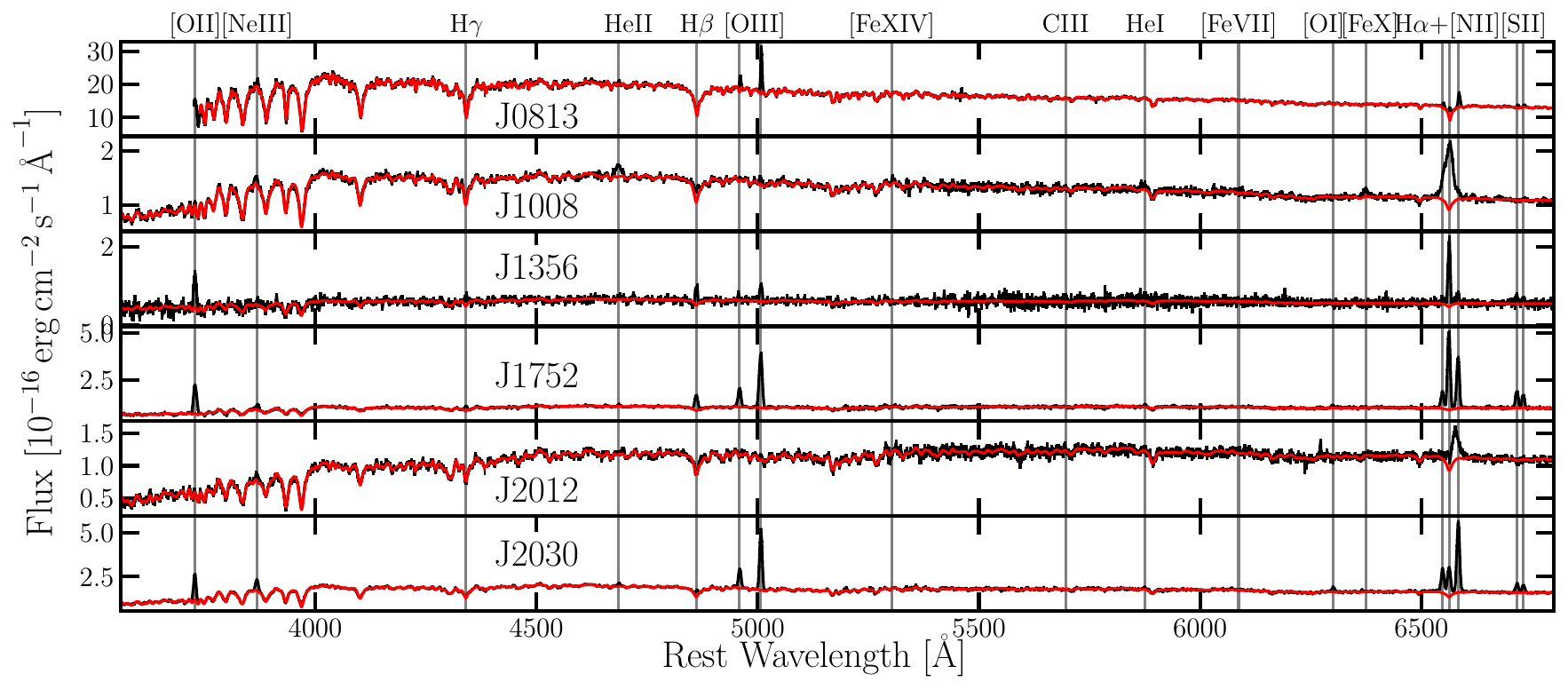}
    \caption{The optical spectra for the radio-selected TDEs in our sample. The observed spectra are shown in black and the best-fit stellar continuum in red (Appendix~\ref{app:optspec}). The Balmer, He\,II, and [Fe\,X] features observed from J1008 and J2012 are associated with the transient and are discussed in \citep[][]{paperII}. }%\todo{Add obs MJDs} }
    \label{fig:opt_specs}
\end{figure*}

In the rest of this paper, our primary goal is to constrain the physical properties that distinguish between radio-selected, optically-bright and optically-selected TDEs. This will rely on observations that are uniformly available for both the radio and optically selected samples; namely, host galaxy photometry and spectroscopy and optical light curves.

We begin by considering the properties of the host galaxies. The properties of host galaxy directly affect the environment in which the transient occurs and the properties of the astrophysical objects that produced the transient; for example, the presence of long-lived ionized emission lines in the optical spectra can suggest the presence of an AGN, or the morphology and color of the host can suggest that the galaxy recently underwent a merger or interaction. The differences in the host galaxies of these radio-selected, optically-detected transients will inform our understanding of the parameters that control radio emission from TDEs.

Each of the following sections considers one or a subset of the host galaxy properties. We begin by describing the relevance of each property, then discuss our methodology for constraining it, and finally compare the radio-selected host properties to those of the optically-selected sample. The host galaxy properties of our sample, constrained as described in the following sections, are summarized in Table~\ref{tab:host}.

\subsection{Black hole and stellar mass} 

SMBHs are central to TDEs and the TDE evolution is certainly affected by the black hole mass. For SMBH masses $M_{\rm BH} \gtrsim 10^8 M_\odot$ the tidal radius $R_T \sim R_*(M_{\rm BH}/M_*)^{2/3}$ is smaller than the Schwarzchild radius $R_s \sim 2 G M_{\rm BH}/c^2$, so any star on an orbit that reaches $R<R_T$ is swallowed whole rather than disrupted. This analysis assumes a non-spinning SMBH: if the SMBH has a high spin, the event horizon is smaller and thus TDEs can occur. SMBH mass may also affect the efficiency with which the TDE accretion disk forms. In some models of TDE evolution, an accretion disk forms when the stellar tidal stream precesses due to Lense–Thirring precession and collides with itself, dissipating energy \citep[][]{Guillochon2015}. The rate of Lense-Thirring precession is $\dot{\Omega} \sim M_{\rm BH}^2$, so these collisions will occur less frequently, if at all, for TDEs by low mass SMBHs  \citep{Lu2020}. This might reduce the high energy (X-ray, UV) emission, if that emission is partly produced during these shocks. Lower precession could also reduce optical emission, if optical emission is produced when outflowing material that becomes unbound during the shocks reprocesses the higher energy emission  \citep{Lu2020}. Radio emission may also be affected; for example, if radio emission is produced from accretion disk winds, as delayed accretion disk formation could produce delayed winds. 

We constrain the SMBH mass using two methods. First, the SMBH mass is tightly connected to the stellar mass \citep{Kormendy2013}. Thus, we begin by simply considering the stellar mass distributions of our galaxies as compared to that of the optically-selected sample. We measure the stellar masses using fits to the UV/optical/IR spectral energy distributions with the \texttt{prospector} SED fitting code \citep[][]{Johnson2021}, following the exact methods of \cite{Hammerstein2022}. The SMBH mass is also tightly connected to the stellar velocity dispersion, which we measure from our high resolution ESI spectra following the same methodology used by \cite{Somalwar2021}.

We can then measure SMBH mass distribution in two ways: (1) using SMBH masses from the host galaxy stellar mass-black hole mass relation from \cite{Yao2023}:
\begin{align}
   {\rm log}&M_{\rm BH,9} = -(1.83\pm0.15) + (1.64\pm 0.27)\times \notag\\
   & {\rm log}\left( \frac{M_\ast}{3\times 10^{10}\,M_\odot}\right);\; {\rm intrinsic\; scatter=0.18},
\end{align}
where $M_{\rm BH,9}$ is the SMBH mass in units of $10^9\,M_\odot$;
(2) for those sources with high-resolution optical spectra, we use SMBH masses measured using the stellar velocity dispersion-black hole mass relation. We adopt the $M_{\rm BH}-\sigma_*$ relation from \cite{Kormendy2013}, to match that used by \cite{Yao2023}:
\begin{gather}
    \log \frac{M_{\rm BH}}{10^9\,M_\odot} = -(0.509\pm0.049)  \nonumber\\ 
    + (4.384 \pm 0.287) \times \log \big(\frac{\sigma_*}{200\,{\rm km\,s}^{-1}} \big).
\end{gather}
The intrinsic scatter in this relation is $0.29$ dex.

\begin{figure*}
    \centering
    \includegraphics[width=\textwidth]{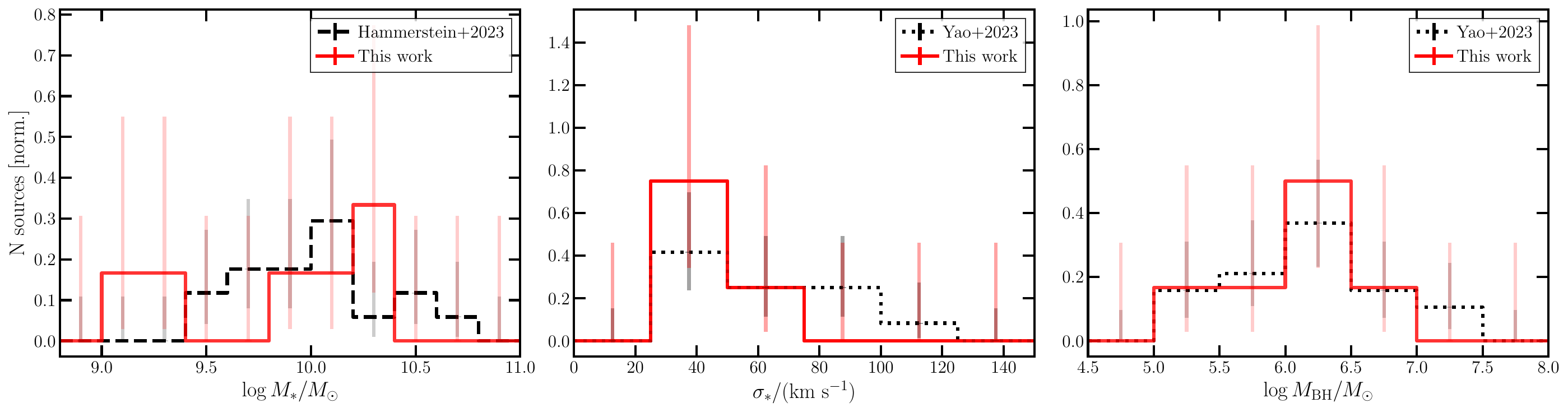}
    \caption{{\it Left panel:} Stellar mass distributions for the radio-selected TDE hosts in this work (red line) as compared to the optically-selected TDE hosts from \cite{Hammerstein2022} (dashed black line). The radio-selected TDEs tentatively prefer lower mass galaxies.  {\it Middle panel:} Stellar velocity dispersion distributions for the radio-selected TDE hosts in this work (red line) as compared to the optically-selected TDE hosts from \cite{Yao2023} (dotted black line). The radio-selected TDEs tentatively prefer lower velocity dispersion galaxies. {\it Right panel:} SMBH mass distributions for the radio-selected TDE hosts in this work (red line) as compared to the optically-selected TDE hosts from \cite{Yao2023} (dotted black line). }
    \label{fig:MBH_comp}
\end{figure*}

We consider the relative black hole masses of the optically-selected and radio-selected samples using three different methods. First, we compare the stellar mass distributions of the two samples, with the caveat that differences in stellar mass can suggest different SMBH masses, or the same SMBH mass but different galactic formation pathways. Then, we compare the stellar velocity dispersion for those objects with stellar velocity measurements, which is a more direct probe of SMBH mass, but is limited by statistics. Finally, we compare the SMBH mass distributions directly, where the masses are inferred from the stellar velocity dispersions and stellar masses.

In Figure~\ref{fig:MBH_comp}, we show histograms of each of these quantities for the radio-selected sample and the optically-selected sample. In the leftmost panel, we show the stellar mass distribution. In the middle panel, we show the stellar velocity dispersion distribution. In the rightmost panel, we show the black hole mass distribution. In all cases, we have chosen the bin size to be larger than the typical measurement uncertainty. While we do not have the statistics to make any definitive claims from these distributions, we can hypothesize possible trends. In both the stellar mass and velocity dispersion panels, the radio-selected TDEs appear to prefer lower values than those of the optically-selected TDEs. This suggests that radio-emitting TDEs tend to occur when the SMBH has a lower mass. Alternatively, there may be differences in the star-formation histories and galactic evolution that lead to radio-emitting TDEs from SMBHs with lower mass and velocity dispersion; such differences will also be considered in Section~\ref{sec:stellarpop} and Section~\ref{sec:discussion}. 

This trend towards lower SMBH masses is not present in the rightmost panel of Figure~\ref{fig:MBH_comp}; however, we urge caution in interpreting this result. While our velocity dispersion measurements agree with \cite{Yao2023}, to which work we are comparing in this figure, our stellar mass measurements follow the methods of \cite{Hammerstein2022} and are consistently larger than those of \cite{Yao2023} (see Figure 24 of \citealp{Yao2023}), which biases our SMBH mass distribution towards higher values. Hence, despite the apparent agreement, we believe there is a tentative trend towards lower SMBH masses for radio-selected TDEs.
   
\subsection{SMBH activity} 

\begin{figure*}
    \centering
    \includegraphics[width=\textwidth]{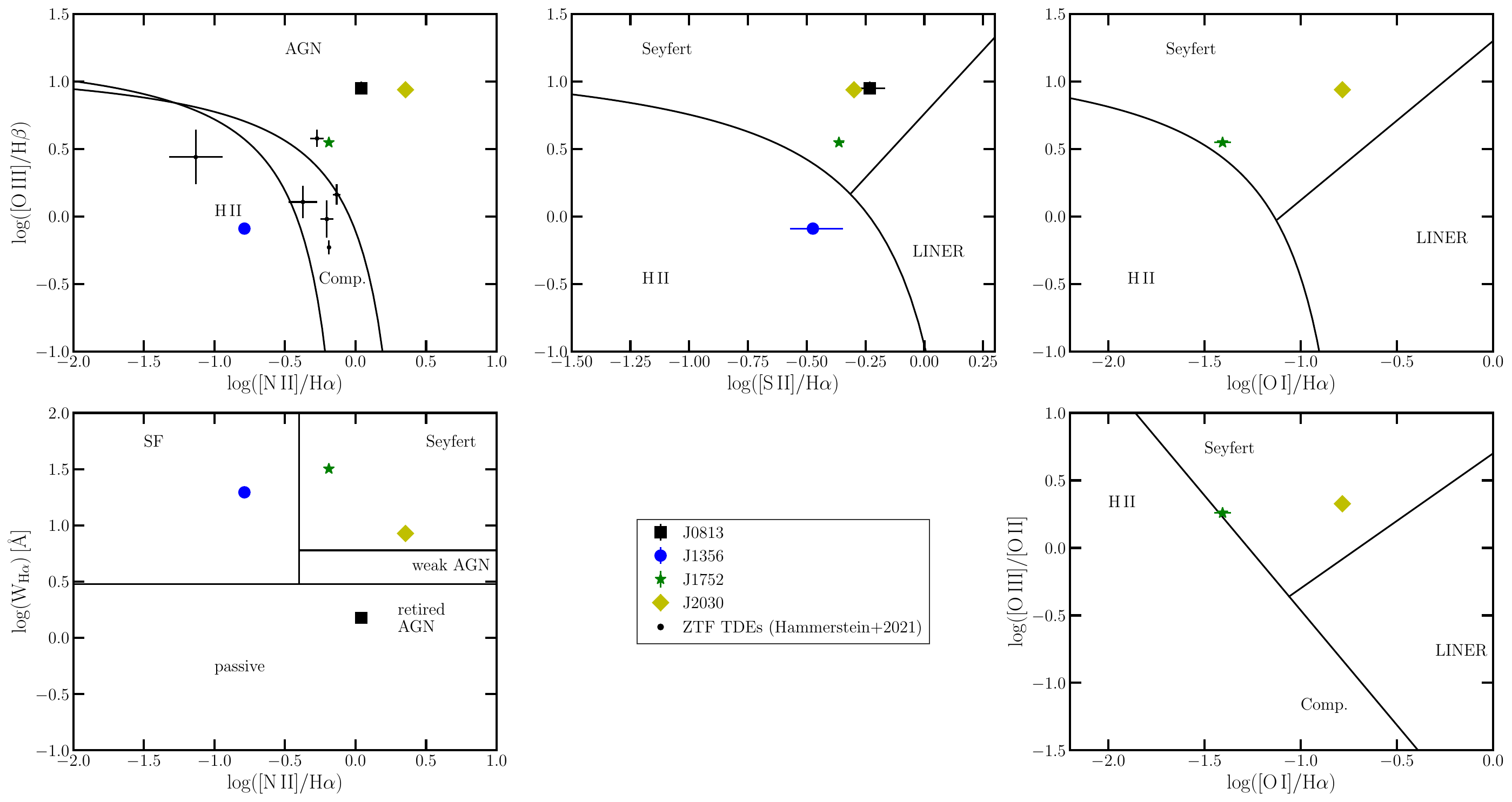}
    \caption{Baldwin, Phillips \& Terlevich (BPT) and W$_{\rm H\alpha}$ versus [N\,II]/${\rm H\alpha}$ (WHAM) diagrams \citep[][]{Baldwin1981, Kewley2006, Cid2011} for those radio- and optically-selected TDEs with detected nebular emission lines. The radio-selected TDEs are shown as colored markers with different shapes, while the optically-selected TDEs are shown as black, thin markers. Note that VT J0813 and VT J1752 are also in the optically-selected sample. The optically-selected TDE hosts with nebular emission appear to favor composite galaxies, whereas the radio-selected TDE hosts favor Seyfert hosts. }
    \label{fig:bpt}
\end{figure*}

TDEs by SMBHs that were actively accreting in the years prior to the event may have substantially altered local environments compared to disruptions by non-accreting SMBHs. AGN tend to have more dust and gas in the nuclei, which will obscure optical emission but can enhance radio and infrared emission. Moreover, a pre-existing accretion disk will alter the properties of the TDE accretion disk; for example, a fossil disk can provide a seed magnetic field that may enable jet launching.  While our selection criteria excludes strong AGN, weak Seyferts or retired AGN are included in our sample.
    
We constrain the SMBH activity in each host using Baldwin, Phillips \& Terlevich (BPT) and W$_{\rm H\alpha}$ versus [N\,II]/${\rm H\alpha}$ (WHAM) diagrams \citep[][]{Baldwin1981, Kewley2006, Cid2011}. We place the host galaxies on BPT and WHAM diagrams using fits to the narrow emission lines in stellar-continuum-subtracted optical spectra. We use the lower resolution LRIS and SDSS spectra for this fit; these are shown in Figure~\ref{fig:opt_specs}. We model the spectral continua and emission lines as described in Appendix~\ref{app:optspec}. The stellar continuum model is shown in red in Figure~\ref{fig:opt_specs}. 

The resulting emission line ratios are shown on the BPT and WHAM diagrams in Figure~\ref{fig:bpt}. In the top left panel, we have overlaid the BPT classifications of the subset of optically-selected TDEs with host galaxy spectral information \citep[][]{Hammerstein2021}. Seven of the nineteen ($(37 \pm 13)\%$, adopting hereafter Poisson frequentist-confidence uncertainties, \citealp{poisson}) optically-selected TDEs considered in that work show strong nebular emission lines, and can thus be placed on a BPT diagram. In contrast, four out of our six events ($(67 \pm 25)\%$) have nebular emission lines. These fractions are consistent within statistical uncertainties, although they may suggest that active/star-forming galaxies tend to host radio-loud AGN.

A larger fraction of our sample lies in the AGN region than found for the optically-selected events: 2/19 ($(11\pm9) \%$) of the optically-selected events are classified as AGN, whereas 3/6 ($(50 \pm 26) \%$) of the radio-selected events are classified as AGN, where we have identified AGN as those host galaxies that are classified as an AGN or Seyfert in the BPT diagrams. We note that VT J0813 is classified as a retired galaxy in the WHAM diagram, suggesting that some of the emission line flux is contributed by an old stellar population. However, the classification of this object as a Seyfert in the BPT diagrams, rather than a LINER, suggests that there may be some emission component resulting from AGN activity: AGN-free, retired galaxies are typically BPT liners \citep[][]{Cid2011}. 4/19 ($(21\pm12) \%$) of the optically-selected events are classified as composite, whereas none of the radio-selected events are classified as such. Finally, 1/19 ($(5\pm8) \%$) of the optically-selected events are classified as star-forming, whereas 1/6 ($(17 \pm 21) \%$) of the radio-selected events are classified as such. These results suggest that radio-selected TDEs may prefer galaxies with more recent or stronger AGN activity, and they may prefer star-forming hosts more than optically-selected TDEs. These trends are not statistically significant, however.

Note that if this AGN activity trend is real (as we do not have the statistics to determine), it may be induced by different selection criteria between the samples -- while both samples exclude strong AGN, the treatment of weak AGN, like those discussed here, is ambiguous. We include those sources that may have had a recent AGN, but are not necessarily still active. The optical selection may be different, leading to some of these trends. The trend with star-forming hosts should be more robust to selection effects.

The presence of recent AGN activity in the hosts of VT J1752, J2030, and likely J0813 may call into question their TDE classification. Low luminosity AGN can produce flares through mechanisms other than stellar disruption, such as accretion disk instabilities, and distinguishing between TDEs and AGN flares is notoriously difficult \citep[e.g.][]{Somalwar2021}. We cannot rule out that these three events are caused by AGN variability. However, in this paper, we have empirically classified these events as TDEs for consistency with the optical selection: all of these events would pass the typical optical TDE selection cuts \citep[such as those used in][]{Yao2023}, and both VT J0813 and VT J1752 are, in fact, included in optical TDE samples. We consider these events to be TDEs for the rest of this work given the consistency with typical optically-selected events and the fact that the AGN are very low luminosity, if on at all. We urge further research on distinguishing between AGN flares and TDEs.

\subsection{Stellar population and star-formation} \label{sec:stellarpop}

\begin{figure*}
    \centering
    \includegraphics[width=0.45\textwidth]{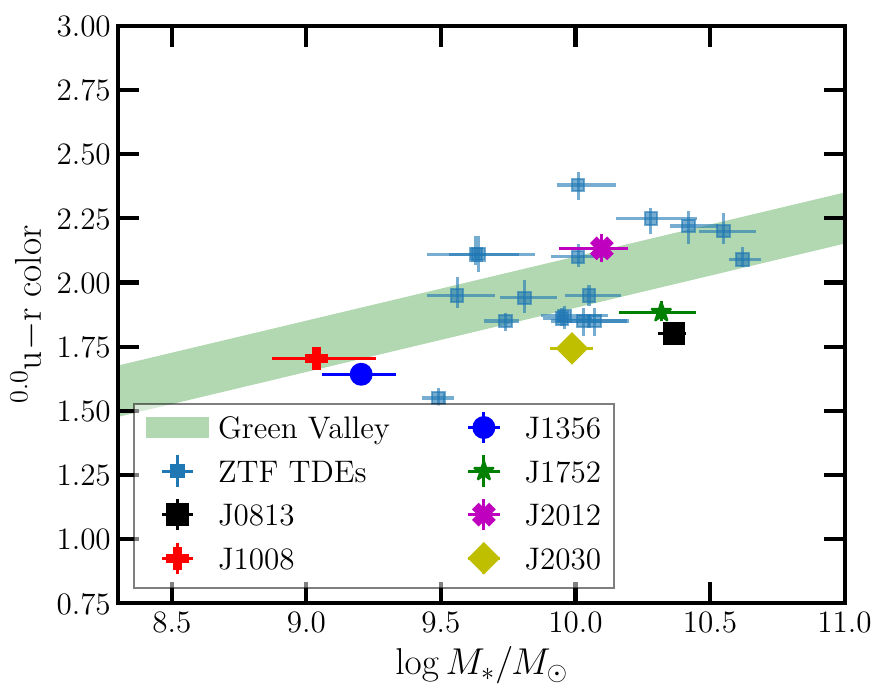}
    \includegraphics[width=0.45\textwidth]{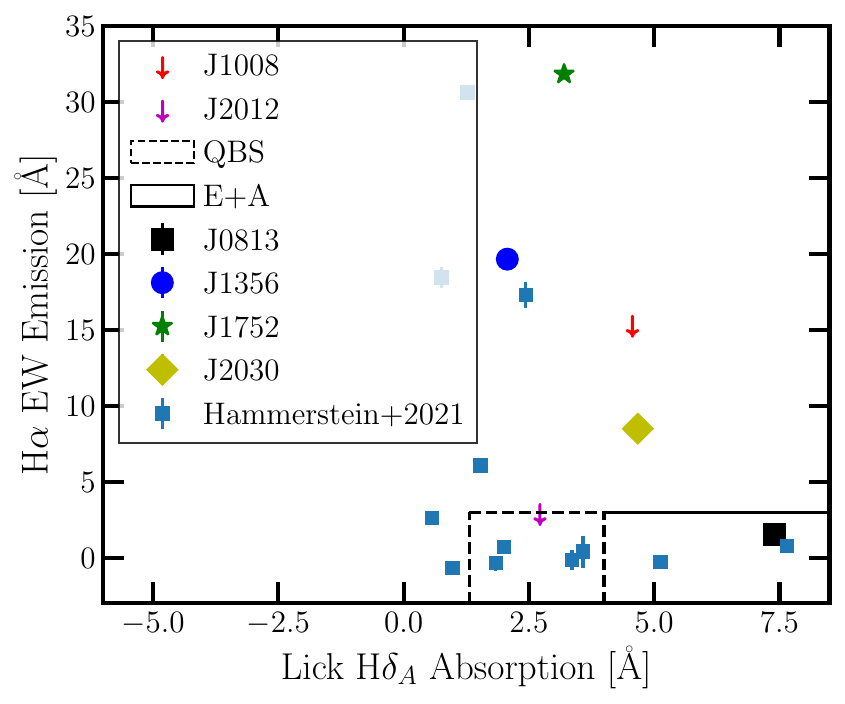}
    \caption{{\it Left panel:} The rest-frame $u-r$ color ($^{0.0}u-r$) of the TDE host galaxies vs the host galaxy stellar mass. The radio-selected sample is shown as colored markers, and the optically-selected sample is shown as light blue squares. The region of this plot occupied by galaxies in the green valley is shown as a green shaded region \citep{Schawinski2014MNRAS.440..889S}. {\it Right panel:} The H$\alpha$ equivalent width versus the Lick H$\delta_A$ absorption for the TDE host galaxies. The different TDEs are formatted the same way as in the {\it left panel}, except upper limits are shown as arrows. }
    \label{fig:stellar_pops}
\end{figure*}

Galactic star-formation histories, and thus stellar populations, are tightly connected to the galaxy evolution. They may reflect processes that can drive stars to the galactic nucleus; for example, mergers can trigger starbursts and may enhance the TDE rate, so recent bursts of star-formation and young ($\lesssim 1$ Gyr) stellar populations may correlate TDE rate. Optically-selected TDE hosts also tend to reside in the green valley, which includes the population of galaxies transitioning from the starforming state (blue cloud) to a non-starforming (red-and-dead) state or vice versa \citep[e.g.][]{greenvalley}. They are well established to predominantly reside in E+A galaxies \citep[][]{French2020SSRv..216...32F, Hammerstein2021}, which are galaxies that have undergone a recent ($\lesssim 1$ Gyr) starburst.

% The star-formation in a galaxy is also connected to its AGN activity history as AGN driven outflows can trigger star-formation. Like mergers, AGN activity will also affect the structure of the nuclear regions of a galaxy, and thus affect the TDE rate. 

% First, we use the H$\alpha$ measurements from the line fits to constrain the star-formation rate in each galaxy. First, we correct the H$\alpha$ flux for host extinction. First, we correct the optical spectra for Galactic extinction using the \cite{Cardelli1989ApJ} extinction curve. Then, assuming Case B recombination with $T=10^4$ K, the color excess E(B-V) given an observed Balmer decrement is \cite{Dominguez2013ApJ}
% \begin{equation}
%     E(B-V) = 1.97 \log \bigg[ \frac{({\rm H}\alpha/{\rm H}\beta)_{\rm obs}}{2.86} \bigg].
% \end{equation}
% We use this extinction combined with the attenuation curve from \cite{Calzetti2000} to calculate a de-reddened H$\alpha$ luminosity and to calculate the star-formation rate. We adopt the H$\alpha$-SFR correlation from \cite{Murphy2011ApJ}: 
% \begin{equation}
%     \bigg( \frac{{\rm SFR}_{{\rm H}\alpha}}{M_\odot {\rm yr}^{-1}} \bigg) = 5.37 \times 10^{-42} \bigg( \frac{L_{\rm H\alpha}}{\text{erg s}^{-1}} \bigg).
% \end{equation}

Because indicators of the presence of an E+A and/or green valley galaxy have been studied in detail for optically-selected TDEs, we focus on those same indicators for our galaxies \citep[][]{Hammerstein2021}. We discuss the star formation histories of the individual VLASS transients in more detail in the companion papers and Appendix~\ref{app:j1356}.

We determine whether a galaxy is in the green valley by comparing its MW extinction and redshift-corrected $u-r$ color, which we denote $^{0.0}u-r$, and stellar mass to those for green valley galaxies, as shown in the left panel of Figure~\ref{fig:stellar_pops}. We adopt the same definition of green valley as \cite{Hammerstein2022} and overplot the $^{0.0}u-r$ colors and stellar masses for their optically-selected sample as blue squares. Qualitatively, all the radio-selected TDE hosts are towards the edges of or outside the green valley, with a preference towards bluer hosts, whereas the optically-selected TDE hosts span it.  Quantitatively, two of the six ($(33\pm25)\%$) radio-selected TDEs fall within the green valley, which is consistent with the eight of the seventeen ($(47\pm 14)\%$) optically-selected TDEs in the same region.

We also constrain the stellar populations using the optical spectra. In particular, we identify E+A galaxies using the Lick H$\delta_A$ absorption index and H$\alpha$ equivalent width, as shown in the right panel of Figure~\ref{fig:stellar_pops}. The E+A region is shown in black solid lines. We have adopted the same E+A galaxy definition as \cite{Hammerstein2022}. We also show the quiescent balmer strong (QBS) region is dashed black lines. Up to 2/6 ($(33\pm25 \%$) of the VLASS TDE hosts are consistent with E+A galaxies, although a quiescent H$\alpha$ luminosity for J1008 is required to confirm that result. This fraction is consistent with the $2/12$ ($(16\pm14 \%$) of the optically-selected TDEs that satisfy the E+A definition.

\subsection{Summary}

Our host galaxy analysis can be summarized as follows:
\begin{enumerate}
    \item Radio-selected, optically-bright TDEs tend to lie at lower stellar masses and SMBH masses than optically-selected TDEs
    \item Radio-selected TDEs occur at a slightly higher rate in galaxies with detectable nebular emission relative to optically-selected TDEs. A larger fraction of these radio-selected events with nebular emission lie in AGN hosts, whereas optically-selected events tend to prefer composite hosts, although this trend may be in part due to different treatments of transients in AGN between the samples.
    \item Radio-selected, optically-bright TDEs occupy green valley galaxies and E+A galaxies at approximately the same rate as the optically-selected TDEs. %Qualitatively, the radio-selected galaxies tend to be bluer than the optically-selected hosts.
\end{enumerate}
Future studies with larger radio-selected TDE samples will test whether these correlations are real, or statistical flukes.

\section{Optical transient broadband emission analysis} \label{sec:optlc}

\begin{figure*}
    \centering
    \includegraphics[width=\textwidth]{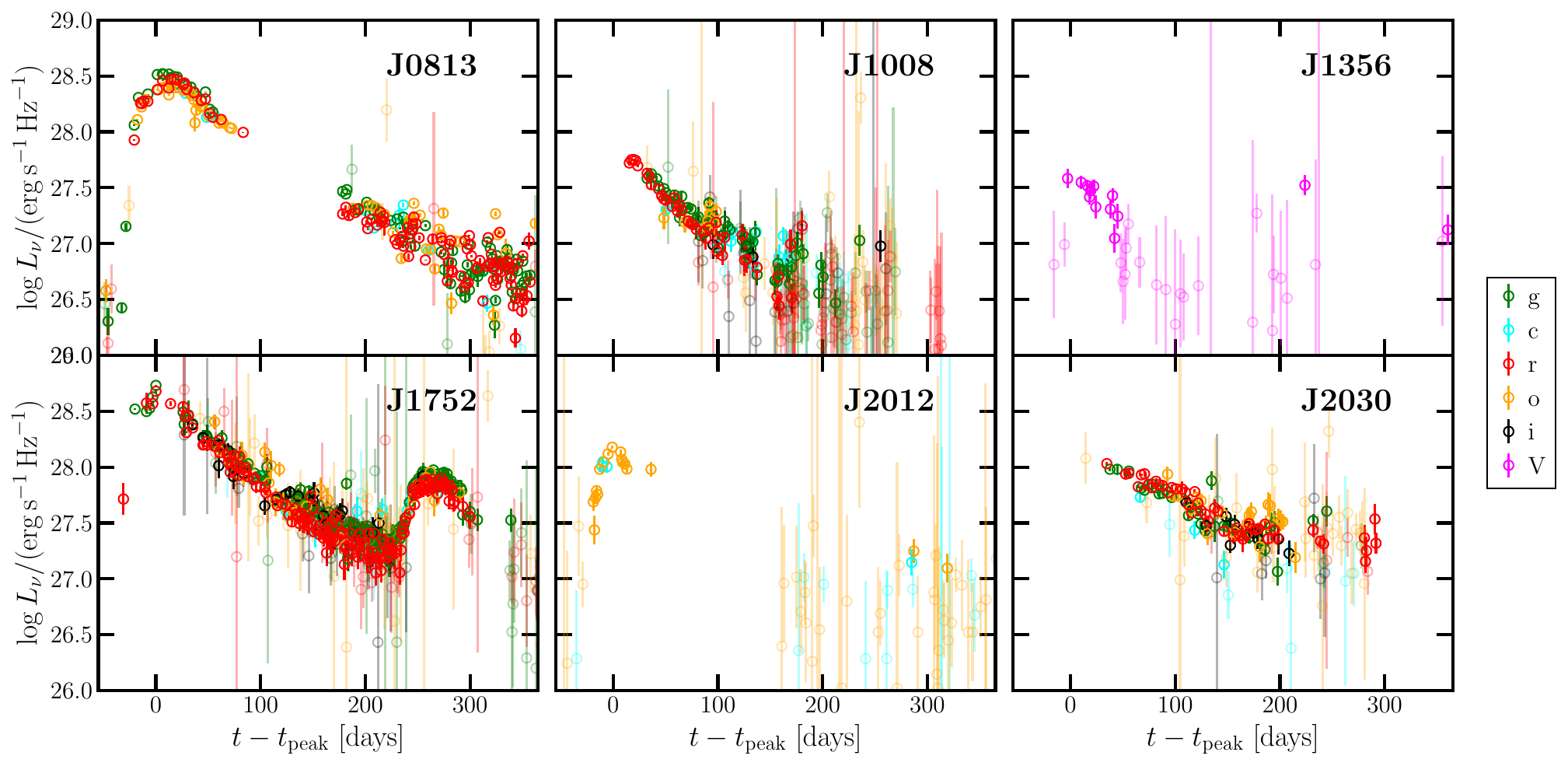}
    \caption{The optical lightcurves for our TDE candidates. Each band is shown in a different color. }
    \label{fig:opt_lc}
\end{figure*}

Informed by the discussion of the host emission from the last section, we begin our analysis of the multiwavelength transient emission. We first consider the optical broadband emission for two reasons: (1) the optical emission from TDEs probes the evolution of the TDE debris and the accretion flow at {\it early} times (${\sim}$months). Later, we will consider the radio emission, which can be delayed and long-lived. (2) A key datum is the MJD on which the TDE began; VLASS alone is not sufficiently high-cadence to constrain this date. The optical light curves set the strongest constraints because they have long baselines and high cadences. In this section, we present the optical light-curves of our events (Figure~\ref{fig:opt_lc}) and perform a basic analysis of their evolution. 

\subsection{Methodology}

{\addtolength{\tabcolsep}{-2pt}   \begin{deluxetable*}{c | cccccc}
\centerwidetable
\tablewidth{10pt}
\tablecaption{Optical light curve parameters \label{tab:optlc}}
\tablehead{ \colhead{Name} & \colhead{$L_{\rm bb}$} & \colhead{$T_{\rm bb}$} & \colhead{$R_{\rm bb}$} & \colhead{$t_{1/2,{\rm decay}}$} & \colhead{$t_{1/2,{\rm rise}}$} & \colhead{Ref.} }
\startdata
 VT J0813 & 44.30 & 4.46 & 14.80 & $45.4^{+1.0}_{-0.9}$ & $24.1^{+1.1}_{-0.9}$ &  \citealp{Yao2023}\cr
 VT J1008 & 42.98 & 4.15 & 14.76 & $23.1^{+1.8}_{-1.2}$ & $11.8^{+1.5}_{-1.3}$ & \citealp{Yao2023} \cr
 VT J1356 & 42.64 & 3.73 & 15.44 & $21.2^{+2.7}_{-2.2}$ & $2.6^{+1.4}_{-0.7}$ & This work \cr
 VT J1752 & 43.81 & 4.10 & 15.28 & $27.4^{+0.7}_{-0.7}$ & $23.1^{+0.9}_{-0.9}$ & \citealp{Yao2023}  \cr
 VT J2012 & 43.07 & 3.93 & 15.26 & $26.2^{+5.4}_{-5.1}$ & $10.2^{+1.5}_{-1.1}$ & This work \cr
 VT J2030 & 43.09 & 3.88 & 15.36 & $66.6^{+2.0}_{-4.4}$ & $15^{+14}_{-10}$ & This work 
\enddata
\centering
\end{deluxetable*}}

TDE optical lightcurves are typically modelled as evolving blackbodies. We aim to constrain four basic parameters of our lightcurves: the peak luminosity $L_{\rm bb}$, the blackbody temperature at the time of the peak luminosity T$_{\rm bb}$, the rise time from half-max-luminosity to max-luminosity $t_{1/2, \rm rise}$, and the decay time from max-luminosity to half-max luminosity $t_{1/2, \rm decay}$. These parameters have already been constrained for J0813, J1008, and J1752 by \cite{Yao2023} and we refer the reader to that work for details.

\begin{figure}
    \centering
    \includegraphics[width=0.45\textwidth]{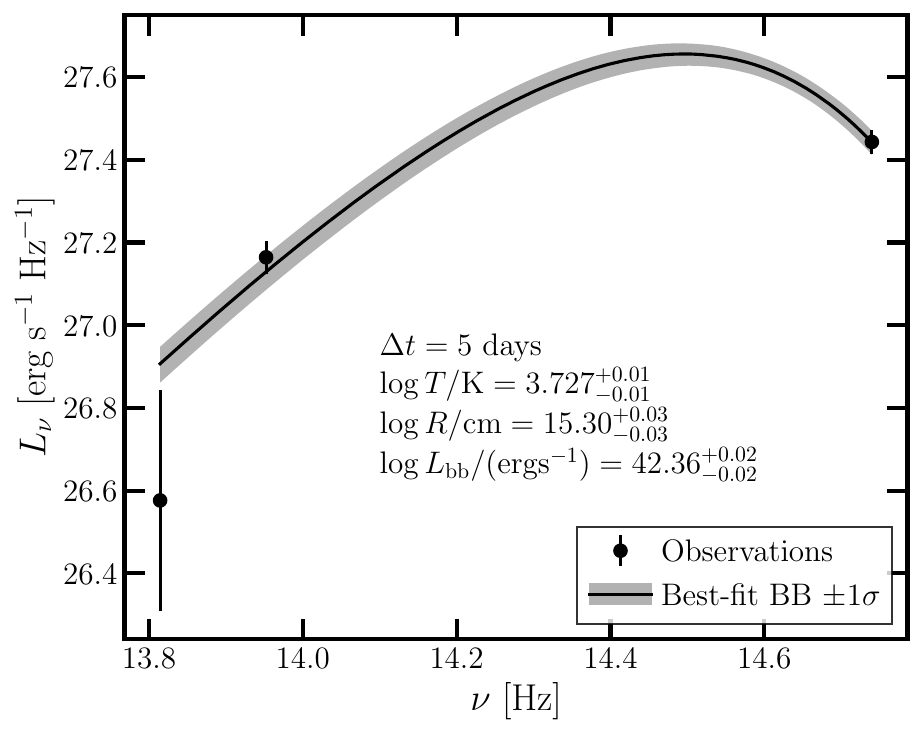}
    \caption{The transient optical-IR SED for VT J1356 near optical peak. The observations are shown as black scatter points. The black line and band shows the best-fit blackbody and $\pm1\sigma$. The best-fit parameters are shown in the figure.}
    \label{fig:J1356_IR}
\end{figure}

In the case of J1356 (described in detail in Appendix~\ref{app:j1356}), we only have one band in the optical, so we do not have enough information to simultaneously fit the blackbody temperature and luminosity from that data alone. Fortuitously, J1356 was observed near the optical peak in the IR, and the resulting IR and optical SED is shown in Figure~\ref{fig:J1356_IR}. We fit this SED to a blackbody and found a best-fit temperature of $\log T/{\rm K} = 3.73^{+0.01}_{-0.01}$. Errors are statistical. We then fit the optical lightcurve to a Gaussian rise and exponential decay model, which is commonly adopted for TDEs. We used the \texttt{dynesty} dynamic nested sampling software.

In the case of J2012, the sparse optical observations preclude any detailed constraints on the optical evolution. We fit the optical lightcurve at $t<1$ year to a Gaussian rise and exponential decay model with a fixed temperature, but urge caution in interpreting the results.

For J2030, there are no optical observations during the rise or peak of the source, so we can only set a limit on the blackbody temperature and luminosity, and we have no constraints on the rise time. We fit the decay to a exponential model with a fixed blackbody temperature and set a lower limit on the peak luminosity based on the luminosity during the first detection of the source.

The adopted optical lightcurve parameters for each source are summarized in Table~\ref{tab:optlc}. 

\subsection{Results}

\begin{figure*}
    \centering
    \includegraphics[width=\textwidth]{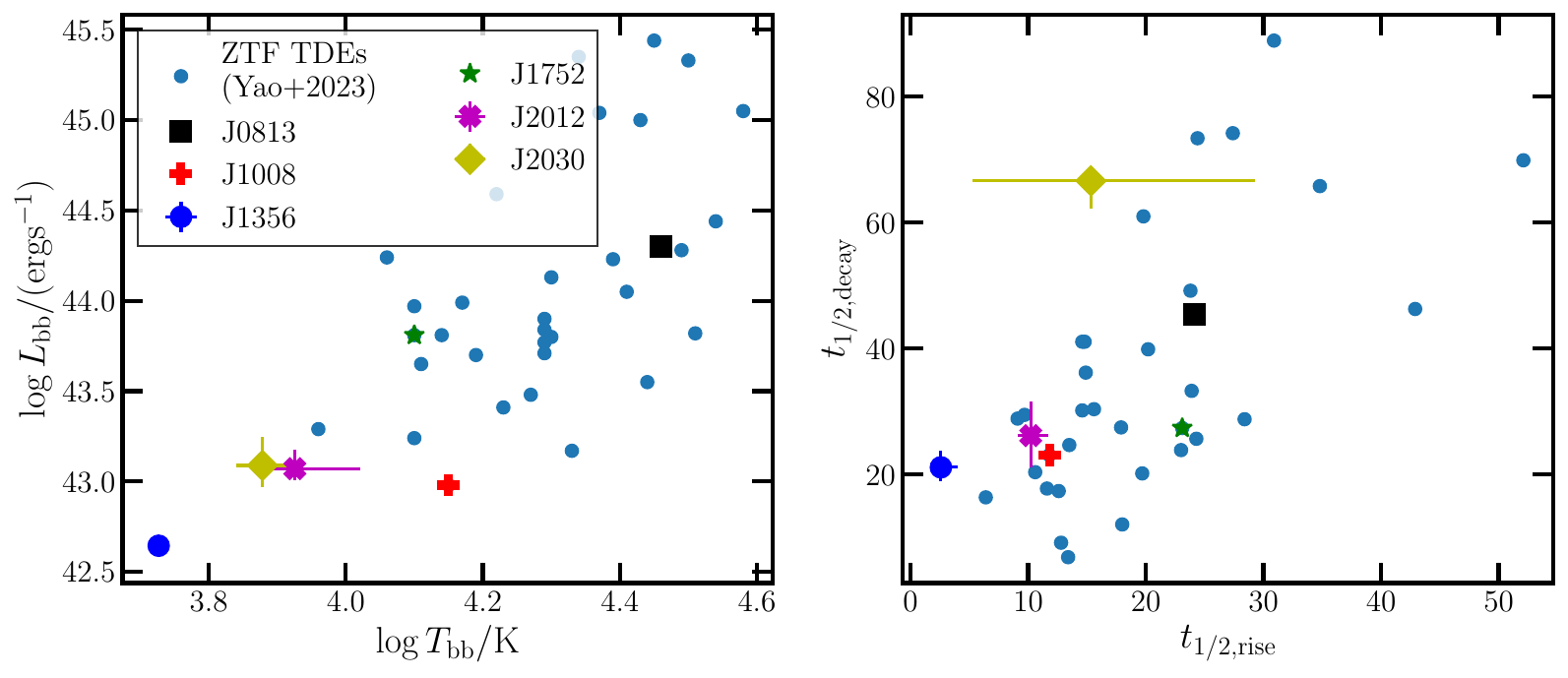}
    \caption{{\it Left panel:} The peak blackbody luminosity vs the blackbody temperature at peak luminosity for the radio-selected TDEs and the optically-selected TDEs, in the same format as other figures. Radio-selected TDEs may have optical lightcurves with lower temperatures and luminosities than those of the typical optically-selected event. {\it Right panel:} The rise versus decay times for the optically- and radio-selected events, in the same format as the {\it left panel}. No obvious trends are present.}
    \label{fig:optlc_pars}
\end{figure*}

In the rest of this section, we compare the optical lightcurve parameters for our radio-selected, optically-detected TDEs to those for the optically-selected TDE sample from \cite{Yao2023}. 

In Figure~\ref{fig:optlc_pars}, we show the distribution of blackbody temperature and luminosity in the left panel and $t_{1/2, {\rm rise}}$ vs $t_{1/2, {\rm decay}}$ in the right panel. We see no obvious correlations in the rise/decay times of the lightcurves. Radio-selected TDE hosts seem to lie towards lower temperature and luminosities than the optically-selected sample, although the bulk of those events at extreme temperatures/luminosities are those without well-sampled data, suggesting this correlation may be an artifact of our analysis methods or data quality. Ignoring those sources (J1356, J2012, J2030), two out of three of the remaining events do lie at low temperatures and luminosities.

\section{Radio emission properties and mechanism} \label{sec:radioana}

\begin{figure*}
    \centering
    \includegraphics[width=\textwidth]{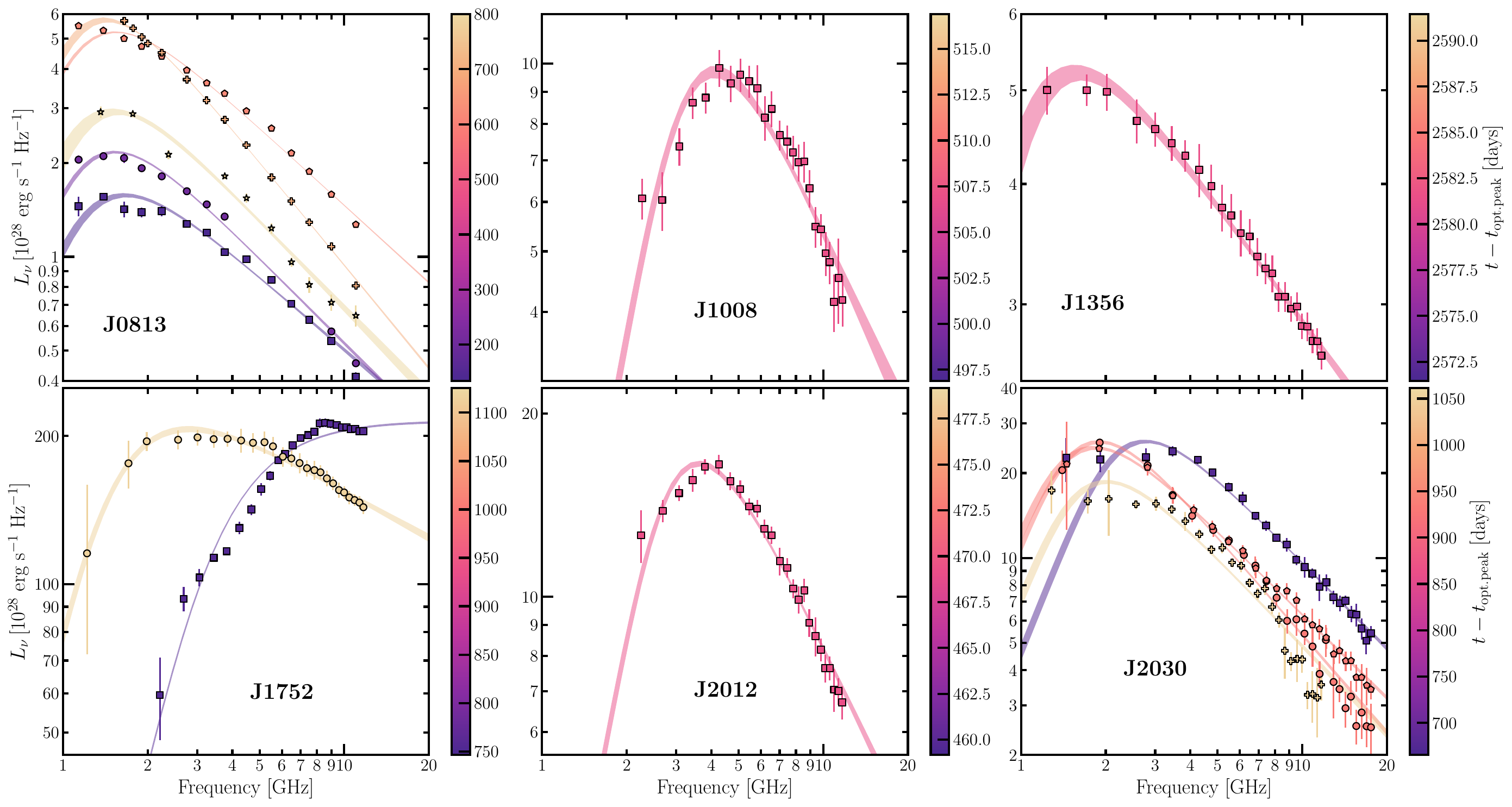}
    \caption{Representative radio SEDs for each VLASS TDE candidate. The observations are shown as scatter points, and the best-fit synchrotron $\pm1\sigma$ models shown as transparent colored bands. The color of the scatter points and fit bands correspond to the time since optical peak of each radio observation. Note that VT J1752 is an example of a source that is very poorly fit by a simple synchrotron model; we adopt this model despite the inconsistent fit for uniformity.}
    \label{fig:radio_seds}
\end{figure*}

Next, we consider the radio emission mechanisms for our sample and compare to published radio follow-up of optically-selected TDEs. A small but growing sample of TDEs have radio detections; the brightest four of which comprise the on-axis, jetted TDE population \citep[][]{jet1,jet2,jet3,jet4,jet5,jet6,jet7,jet8,jet9,Somalwar2023ApJ...945..142S}. Most TDEs with radio detections have non-relativistic, wide-angle outflows. The non-relativistic TDEs were largely selected in the optical and followed up in the radio, with a few exceptions. In the following sections, we first consider the radio lightcurves of our events in the context of published TDE radio lightcurves. Then, we constrain the physical parameters of the emitting region and compare these to those of optically-selected, radio-detected TDEs.

\begin{figure*}
    \centering
    \includegraphics[width=\textwidth]{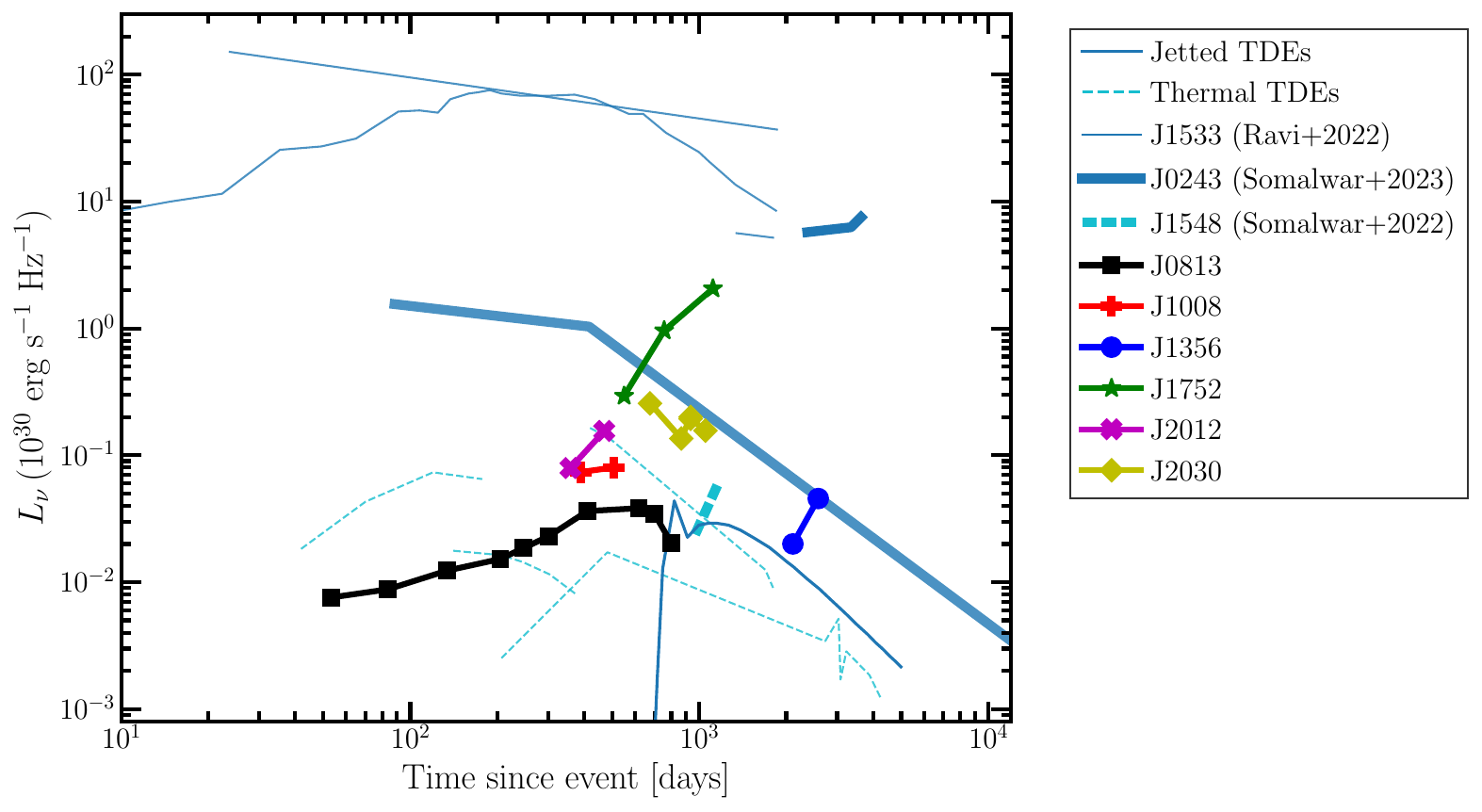}
    \caption{GHz radio lightcurves for a representative sample of radio-detected TDEs. The solid, blue lines show example jetted TDEs and the dashed, cyan lines show the non-relativistic TDE sample. The thick lines correspond to objects that were found in untargeted radio searches, whereas the thin lines correspond to objects that were selected in other bands \citep[see][and references therein]{Somalwar2023ApJ...945..142S}.}
    \label{fig:radio_lc}
\end{figure*}

\subsection{The radio lightcurves of TDEs}
First, we show the radio lightcurves for our TDEs and representative sub-sample of all radio-detected TDEs in Figure~\ref{fig:radio_lc}. The jetted and non-jetted TDEs are denoted with different line styles and colors. The radio-selected TDEs all have thicker lines than those selected in other bands. The TDEs presented in this paper are demarcated with unique colors and markers. 

There are a number of key takeaways from this plot. From the non-radio-selected TDEs alone, there was an apparent bifurcation of the events into those with luminosities $\gtrsim 10^{31}$ erg s$^{-1}$ Hz$^{-1}$ and those with luminosities $\lesssim 10^{29}$ erg s$^{-1}$ Hz$^{-1}$. This split approximately mapped onto the type of outflow: relativistic, collimated jet, or non-relativistic, wide-angle outflow. With the radio-selected TDEs, we have begun spanning the gap between these two outflow classes. For example, we see evidence for jetted sources at luminosities $\lesssim 10^{30}$ erg s$^{-1}$ Hz$^{-1}$, like J1533 from \cite{Ravi2021} and J2030 from this work. The radio-selected TDEs alone span a luminosity range $10^{28-31}$ erg s$^{-1}$ Hz$^{-1}$, with no clear division into separate classes.

The fact that the radio-selected sample has more luminous emission may be a selection effect - radio surveys are biased towards brighter radio sources. The fact that optical surveys do not see these bright sources results from the trends discussed in previous sections; in particular, it arises naturally from the optical faintness of radio-selected TDEs.

We also see a wide range of lightcurve shapes. Until recently, TDE radio lightcurves were expected to follow gamma-ray burst models, with a power law rise and decay. This paradigm shifted with the discovery of $\gtrsim 5$-year-old radio-selected TDEs \citep[e.g.][]{Ravi2021, Somalwar2023ApJ...945..142S} and the discovery of radio rebrightening and other late-time radio emission from optically-selected TDEs \citep[e.g.][]{Horesh2021}. The radio TDEs presented in this work extend the range of radio lightcurves. We see a wide range of light curve shapes, including multiple examples that show late-time rise or rebrightenings. Some of the rises and rebrightenings occur remarkably long after the initial TDE. VT J2030 declines and then brightens ${\sim}3$ years post-optical flare, while VT J1752 is still rising at that time. VT J1356 is still rising after ${\sim}8$ years, making this one of the oldest known radio-emitting TDEs, and the first known in a non-active galaxy. We also see examples like VT J1008 and VT J2012, which show similar multiwavelength properties and are at similar epochs post-TDE, but the 3 GHz luminosities are rising at significantly different rates.

Of course, these single frequency lightcurves cannot provide a complete view of the physical properties of the radio-emitting outflows. Instead, we require multi-frequency radio SEDs (i.e., near-simultaneous observations at multiple frequencies, spanning a few GHz range) spanning  multiple epochs, which we can fit to radio outflow models. Hence, for the rest of this section, we analyze the radio SEDs of our events and briefly compare them to published results for optically-selected, radio-detected TDEs.

\subsection{Analysis of the TDE radio spectral energy distributions}
 
We modelled the SED of each of the TDEs presented in this work uniformly. A detailed model description is provided in Appendix~\ref{app:synch}, which we briefly summarize here. We assume all TDE radio emission is produced by the synchrotron mechanism. We model each radio SED as a spherical outflow of radius $R$ expanding into a medium with uniform density and magnetic field $B$. We assume a single power-law electron population with density $N_0$, minimum Lorentz factor $\gamma = \gamma_{\rm min}$ and spectral index $p$:
\begin{equation}
    N(\gamma) \propto \gamma^{-p}, \gamma_{\rm min}< \gamma.
\end{equation}
With these definitions, the synchrotron flux density can be calculated as a function of frequency, with free parameters $R$, $B$, $n$, $N_0$, and $\gamma_{\rm min}$. We assume equipartition with $\epsilon_{\rm e} = 0.1$ and $\epsilon_{\rm B} = 0.1$, as is standard in TDE modelling. This assumption allows us to eliminate one parameter from our fit. We fit the synchrotron flux density to each radio SED independently (i.e., assuming no model for time evolution) using the \texttt{dynesty} dynamic nested sampler.

In the case of J1752, the radio emission model required with two components. Our methodology was otherwise identical to that used when modelling the rest of the sample.

The best-fit synchrotron parameters for each source at each epoch are summarized in Table~\ref{tab:radio_sample}. We additionally include the total energy in the outflow $E$, the equivalent ejecta mass $M_{\rm ej} = 2E/c^2$, and the average $\beta$ factor of the ejecta ($\beta = v/c$, for velocity $v$) assuming the first optical detection is the launch date of the outflow. For illustration, we show a representative sample of the radio SEDs for the six TDEs published in this paper in Figure~\ref{fig:radio_seds}, with the fits overlaid. 

\subsection{Comparison to optically-selected, radio-detected TDEs}

In the rest of this section, we discuss the radio emission observed from both our sample and place it in the context of the general radio-detected TDE population. As a comparison, we primarily rely on \cite{cendes_radiooptsamp}, who presented radio follow-up of 23 optically-selected TDEs, although without a well-defined selection function. Radio emission was detected from $15$ of the sources, although the transient nature of the emission could only be confirmed for $9$ sources. Thus at least $40\%$ of the optically-selected TDEs that were considered were radio emitting.

\cite{cendes_radiooptsamp} obtained multi-frequency SEDs for their sample and constrained the radii, magnetic field, and energies using methods similar to ours although using a broken power-law fit to measure the peak flux/frequency of the radio SEDs; we refer the reader to that work for details. In the rest of this section we will compare these physical parameters between our sample and that in \cite{cendes_radiooptsamp}, while keeping in mind that the differences is SED fitting methods may result in small offsets in the measured outflow physical parameters, 

\begin{figure*}
    \centering
    \includegraphics[width=\textwidth]{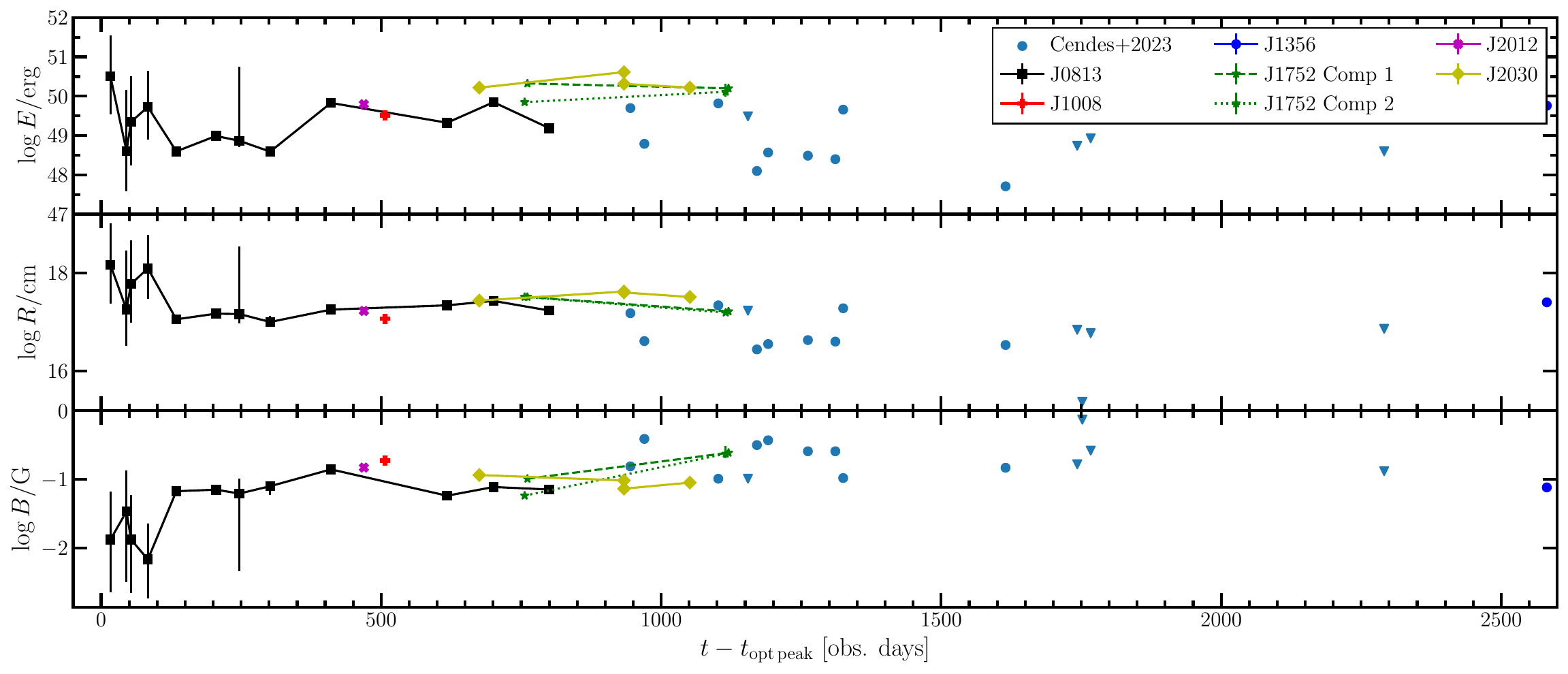}
    \caption{The evolution of the equipartition energies ({\it top}), radii ({\it middle}, and magnetic fields ({\it bottom} for the TDEs in our sample (colored markers) compared to those in the \cite{cendes_radiooptsamp} sample (blue circles and limits). The objects from the \cite{cendes_radiooptsamp} with unconstrained peak flux densities/frequencies are shown as limits. Our events tend to occupy a higher energy, lower magnetic field, large radius state than the objects in the \cite{cendes_radiooptsamp} sample. }
    \label{fig:radio_pars}
\end{figure*}

In Figure~\ref{fig:radio_pars}, we show the radius, energy, and magnetic field evolution as a function of time. We do not adopt the energies reported in \cite{cendes_radiooptsamp}, but instead adopt $E = \frac{B^2}{8 \pi \epsilon_B} 4\pi R^3 f_V$, where $f_V$ is the volume filling factor. The clearest discrepancy between our sources and the sources from \cite{cendes_radiooptsamp} is that the latter turn on in the radio later than our sources. Our sources also tend to reside at the upper end of the energy and radius range observed by \cite{cendes_radiooptsamp}, and the lower end of the magnetic field range.

These differences between our sample and the \cite{cendes_radiooptsamp} sample show that radio and optical searches for TDEs are currently finding fundamentally different events, even when only considering those TDEs that emit in both bands. Our selection criteria, combined with the flux-limited nature of VLASS and the timeline between the start of ZTF and VLASS E1/2 will tend to produce events that brighten within ${\sim}3$ years post-optical flare and that are intrinsically luminous. As we have argued, these sources tend to have faint optical flares, so they will not be as dominant in the \cite{cendes_radiooptsamp}, which is based on optically-discovered TDEs and so will be biased towards objects with brighter optical flares. The \cite{cendes_radiooptsamp} is also based on deep radio follow-up, and thus is sensitive to events with fainter radio emission than is detectable by VLASS.

\section{Discussion} \label{sec:discussion}

In the following sections, we briefly comment on physical explanations for the tentative trends observed in this work and we constrain the rate of radio-emitting, optically-bright TDEs. First, we recapitulate the results of the following sections.

\begin{itemize}
    \item The radio-selected TDEs show a tentative preference for lower stellar mass and velocity dispersion galaxies, suggesting that radio-selected TDEs may prefer low SMBH masses.
    \item Radio-selected TDEs are more likely to occur in galaxies with nebular emission lines than optically-selected TDEs. The radio-selected TDE hosts with nebular emission lines are predominantly Seyferts, whereas the optically-selected TDE hosts with nebular emission lines are predominantly composite galaxies.
    \item Radio-selected TDE hosts occupy E+A galaxies and green valley galaxies at a consistent rate to that of optically-selected TDE hosts. %Qualitatively, the radio-selected TDE hosts tend to prefer bluer galaxies relative to the optically-selected TDE hosts.
    \item The optical light-curves for the radio-selected TDEs tend to peak at lower blackbody temperatures and luminosities than those of the optically-selected TDEs, although this difference could be caused by the difference in available optical datasets for these sources.
    \item TDEs show a wide-range of radio emission. In our sample, alone, we have TDEs that, at ${\sim}3$-years post-first-detection, are both rising and fading at 3 GHz. We have TDEs with strongly non-monotonic evolution, and some with brightening 3 GHz emission almost a decade post-event. The radio luminosities of the radio-selected TDEs alone span the range $10^{28-31}$ erg s$^{-1}$ Hz$^{-1}$, with no obvious division into subclasses. 
    \item Our radio-selected TDEs tend to have emitting regions with larger energies and radii and smaller magnetic fields than optically-selected/radio-detected TDEs, despite the fact that the radio-selected TDE radio emission turns on at earlier times.
\end{itemize}

\subsection{Physical explanations for the observed differences between radio- and optically-selected TDEs}

With these observations in mind, we briefly hypothesize physical explanations for the observed trends. A comprehensive review of possible models is beyond the scope of this paper.

If we adopt the model for TDE evolution where radio emission from TDEs may be produced when a wind or jet launched from near an accretion disk, it is possible to unify the observed trends. First, we consider the low SMBH masses. TDEs with lower black hole masses spend more time in a near- to super-Eddington state \citep[][]{Wu2018MNRAS.478.3016W}. Super-Eddington accretion disks have a puffy structure, with significant unbound material, that may be conducive to launching winds and/or aiding in collimating jets  \citep[][]{Curd2023MNRAS.519.2812C}. Hence, one might expect that radio emitting TDEs tend to occur when the SMBH mass is lower because of the puffier accretion disk structure.

Of course, the wind and/or jet cannot necessarily be detected unless it shocks against the circumnuclear material (CNM). Then, events with more material in the vicinity of the SMBH will be more likely to produce detectable radio emission. It is well established that AGN tend to have a significant CNM relative to completely quiescent galaxies  \citep[][]{Jiang2021ApJ...911...31J}, so a TDE in a galaxy with a weak or retired AGN may be more likely to produce synchrotron-emitting shocks. This would then explain the prevalence of galaxies with strong nebular emission in our sample.

The faint and cool optical flares may also be explained by the enhanced circumnuclear material, combined with the low black hole masses. Lower black hole masses produce fainter optical flares \citep{mummery_lc}. Moreover, the presence of enhanced circumnuclear material will tend to obscure emission from the nucleus. Then, before correcting for this obscuration, any detected optical flare will appear fainter and cooler. If we had higher quality optical observations, it is plausible that we could constrain the amount of extinction caused by the CNM, although this is not possible with the current data. Another possible method of constraining increased CNM in radio-selected TDE hosts is through MIR studies: we would have direct evidence for the presence of enhanced circumnuclear dust if radio-selected TDEs are more likely than optically-selected events to have a MIR flare, which in TDEs is associated with thermally-emitting dust in the vicinity of the SMBH.

Thus, by invoking super-Eddington accretion around low mass SMBHs and enhanced circumnuclear material, we can reconcile most of our observations. Future samples of radio-selected TDEs with larger statistics and improved multiwavelength coverage will be able to probe these possibilities.

\subsection{The rate of radio-emitting, optically-bright TDEs}

One of the key benefits of performing a untargeted search for optically- and radio-detected TDEs is that we can set limits on the rate of such events. Hence, we conclude our discussion by considering the rate of GHz-emitting, optically-bright TDEs on ${\sim}3$ year timescales. 

Because the GHz lightcurve evolution of TDEs is poorly understood, we cannot robustly estimate the incompleteness of VLASS to radio-emitting, optically-detected TDEs. Instead, we simply set an lower limit, which acknowledges that fact that there may be some optically-detected, radio-bright TDEs that did not make it into our sample because of, e.g., the cadence of sensitivity of VLASS. Of the thirty-three ZTF TDEs presented in \cite{Yao2023}, three are in our sample. Hence, with $90\%$ confidence, $>3\%$ of optically-selected TDEs are radio-emitting. The integrated volumetric TDE rate from \cite{Yao2023} is $2.9^{+0.6}_{-1.3} \times 10^{-7}$ Mpc$^{-3}$ yr$^{-1}$, leading to a volumetric radio-emitting, optically-bright TDE rate of $\gtrsim 10$ Gpc$^{-3}$ yr$^{-1}$. 

We defer discussion of the luminosity function of radio-selected TDEs, as well as predictions for future surveys, to future work on the full VLASS-selected TDE sample.

% Of these three radio-detected events, two show evidence for relativistic jets \citep[][]{paperIII}. Thus, $>2\%$ of optically-selected TDEs may have collimated jets, or $\gtrsim 6$ Gpc$^{-3}$ yr$^{-1}$. This result is a factor of ${\sim}600$ off from the jetted TDE rate inferred from the known on-axis events: $0.02^{+0.04}_{-0.01}$ Gpc$^{-3}$ yr$^{-1}$ \citep[][]{Andreoni2022Natur.612..430A}. This implies that the known jetted TDEs are only the tip of the iceberg: there is a large population of weak and low energy or off-axis jets that can be produced by TDEs. In fact, given that $2/3$ of the optically-detected, radio-loud TDEs have evidence for a collimated jets, it is plausible that most ${\sim}$year-timescale radio-emitting TDEs produce a weak jet. We explore these possibilities extensively in \citep[][]{paperIII}.

\section{Conclusions}

We have presented the first sample of radio-selected TDEs. We selected six radio transients in the VLASS with multiwavelength emission consistent with TDEs; in particular, we require the detection of an optical counterpart. We have compared the properties of these events to optically-selected TDE samples. We tentatively suggest that radio-selected, optically-bright TDEs occur at a higher rate in galaxies with low stellar/black hole masses. They also tend to have cooler and fainter optical emission that optically-selected TDEs. We compare to the results of radio follow-up campaigns of optical TDE samples and find slightly larger energies and radii in our sample, as well as earlier outflow launch times relative to the initial TDE. We constrain the rate of radio-emitting, optically-bright TDEs to be $>3\%$ of the optical TDE rate, or $\gtrsim 10$ Gpc$^{-3}$ yr$^{-1}$.

In future work, we will use VLASS to extend this sample to include those events without multiwavelength counterparts. VLASS, of course, is limited by the observation cadence: it will only be able to identify TDEs with radio emission that lasts for ${\sim}3$ years. By combining VLASS with past radio surveys like FIRST or NVSS, we can probe long timescales. Current and planned radio surveys and instruments, including ASKAP \citep[][]{askap}, the ngVLA \citep[][]{ngvla} and the DSA 2000 \citep[][]{dsa2000}, will probe different timescales. The combined radio TDE samples from these surveys will revolutionize our understanding of the landscape of TDE radio emission.

\begin{acknowledgments}
Based on observations obtained with the Samuel Oschin Telescope 48-inch and the 60-inch Telescope at the Palomar Observatory as part of the Zwicky Transient Facility project. ZTF is supported by the National Science Foundation under Grants No. AST-1440341 and AST-2034437 and a collaboration including current partners Caltech, IPAC, the Weizmann Institute of Science, the Oskar Klein Center at Stockholm University, the University of Maryland, Deutsches Elektronen-Synchrotron and Humboldt University, the TANGO Consortium of Taiwan, the University of Wisconsin at Milwaukee, Trinity College Dublin, Lawrence Livermore National Laboratories, IN2P3, University of Warwick, Ruhr University Bochum, Northwestern University and former partners the University of Washington, Los Alamos National Laboratories, and Lawrence Berkeley National Laboratories. Operations are conducted by COO, IPAC, and UW.

The ZTF forced-photometry service was funded under the Heising-Simons Foundation grant \#12540303 (PI: Graham).

Some of the data presented herein were obtained at the W. M. Keck Observatory, which is operated as a scientific partnership among the California Institute of Technology, the University of California and the National Aeronautics and Space Administration. The Observatory was made possible by the generous financial support of the W. M. Keck Foundation.

The authors wish to recognize and acknowledge the very significant cultural role and reverence that the summit of Maunakea has always had within the indigenous Hawaiian community.  We are most fortunate to have the opportunity to conduct observations from this mountain.

This material is based upon work supported by the National Science Foundation Graduate Research Fellowship under Grant No. DGE‐1745301.

The Legacy Surveys consist of three individual and complementary projects: the Dark Energy Camera Legacy Survey (DECaLS; Proposal ID \#2014B-0404; PIs: David Schlegel and Arjun Dey), the Beijing-Arizona Sky Survey (BASS; NOAO Prop. ID \#2015A-0801; PIs: Zhou Xu and Xiaohui Fan), and the Mayall z-band Legacy Survey (MzLS; Prop. ID \#2016A-0453; PI: Arjun Dey). DECaLS, BASS and MzLS together include data obtained, respectively, at the Blanco telescope, Cerro Tololo Inter-American Observatory, NSF’s NOIRLab; the Bok telescope, Steward Observatory, University of Arizona; and the Mayall telescope, Kitt Peak National Observatory, NOIRLab. Pipeline processing and analyses of the data were supported by NOIRLab and the Lawrence Berkeley National Laboratory (LBNL). The Legacy Surveys project is honored to be permitted to conduct astronomical research on Iolkam Du’ag (Kitt Peak), a mountain with particular significance to the Tohono O’odham Nation.

NOIRLab is operated by the Association of Universities for Research in Astronomy (AURA) under a cooperative agreement with the National Science Foundation. LBNL is managed by the Regents of the University of California under contract to the U.S. Department of Energy.

This project used data obtained with the Dark Energy Camera (DECam), which was constructed by the Dark Energy Survey (DES) collaboration. Funding for the DES Projects has been provided by the U.S. Department of Energy, the U.S. National Science Foundation, the Ministry of Science and Education of Spain, the Science and Technology Facilities Council of the United Kingdom, the Higher Education Funding Council for England, the National Center for Supercomputing Applications at the University of Illinois at Urbana-Champaign, the Kavli Institute of Cosmological Physics at the University of Chicago, Center for Cosmology and Astro-Particle Physics at the Ohio State University, the Mitchell Institute for Fundamental Physics and Astronomy at Texas A\&M University, Financiadora de Estudos e Projetos, Fundacao Carlos Chagas Filho de Amparo, Financiadora de Estudos e Projetos, Fundacao Carlos Chagas Filho de Amparo a Pesquisa do Estado do Rio de Janeiro, Conselho Nacional de Desenvolvimento Cientifico e Tecnologico and the Ministerio da Ciencia, Tecnologia e Inovacao, the Deutsche Forschungsgemeinschaft and the Collaborating Institutions in the Dark Energy Survey. The Collaborating Institutions are Argonne National Laboratory, the University of California at Santa Cruz, the University of Cambridge, Centro de Investigaciones Energeticas, Medioambientales y Tecnologicas-Madrid, the University of Chicago, University College London, the DES-Brazil Consortium, the University of Edinburgh, the Eidgenossische Technische Hochschule (ETH) Zurich, Fermi National Accelerator Laboratory, the University of Illinois at Urbana-Champaign, the Institut de Ciencies de l’Espai (IEEC/CSIC), the Institut de Fisica d’Altes Energies, Lawrence Berkeley National Laboratory, the Ludwig Maximilians Universitat Munchen and the associated Excellence Cluster Universe, the University of Michigan, NSF’s NOIRLab, the University of Nottingham, the Ohio State University, the University of Pennsylvania, the University of Portsmouth, SLAC National Accelerator Laboratory, Stanford University, the University of Sussex, and Texas A\&M University.

BASS is a key project of the Telescope Access Program (TAP), which has been funded by the National Astronomical Observatories of China, the Chinese Academy of Sciences (the Strategic Priority Research Program “The Emergence of Cosmological Structures” Grant \# XDB09000000), and the Special Fund for Astronomy from the Ministry of Finance. The BASS is also supported by the External Cooperation Program of Chinese Academy of Sciences (Grant \# 114A11KYSB20160057), and Chinese National Natural Science Foundation (Grant \# 12120101003, \# 11433005).

The Legacy Survey team makes use of data products from the Near-Earth Object Wide-field Infrared Survey Explorer (NEOWISE), which is a project of the Jet Propulsion Laboratory/California Institute of Technology. NEOWISE is funded by the National Aeronautics and Space Administration.

The Legacy Surveys imaging of the DESI footprint is supported by the Director, Office of Science, Office of High Energy Physics of the U.S. Department of Energy under Contract No. DE-AC02-05CH1123, by the National Energy Research Scientific Computing Center, a DOE Office of Science User Facility under the same contract; and by the U.S. National Science Foundation, Division of Astronomical Sciences under Contract No. AST-0950945 to NOAO.
\end{acknowledgments}

\bibliography{sample631}{}
\bibliographystyle{aasjournal}

\begin{appendix} 
\section{The multi-wavelength properties of VT J1356} \label{app:j1356}

\begin{figure*}
    \centering
    \includegraphics[width=\textwidth]{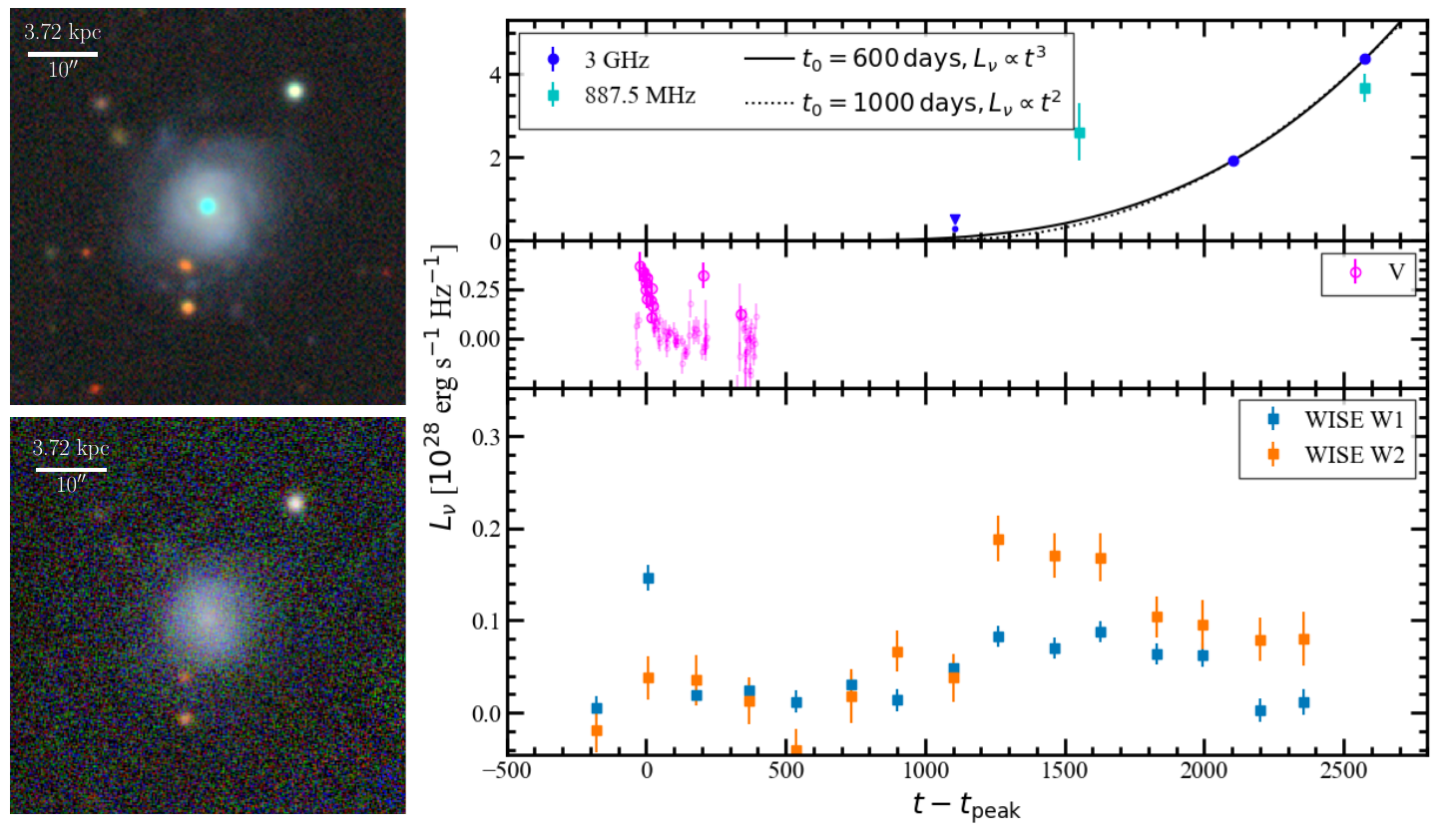}
    \caption{Summary of the properties of VT J1356. The {\it bottom left} panel shows a PANSTARRS image of the host galaxy before the flare. The {\it top left} panel shows a Legacy survey \citep[][]{legacy} image of the host galaxy and transient during the flare. The transient is visible as a blue nucleus in the galaxy. The {\it right} panels summarize the transient emission associated with the event. The {\it top right} panel shows the 3 GHz and 887.5 MHz radio lightcurves. Example power law fits to the 3 GHz lightcurve are overlaid in black. The {\it middle right} panel shows the ASASSN lightcurve of this source. The {\it bottom right} panel shows the WISE infrared lightcurve of this source \citep[][]{wise}. }
    \label{fig:J1356}
\end{figure*}

In this appendix, we discuss the multiwavelength properties of VT J1356 in detail because this source is not discussed in either of the companion papers. 

\subsection{Discovery and host galaxy}
We first identified VT J1356 as a radio flare with an unusual MIR counterpart in data from the WISE survey, shown in the right panel of Figure~\ref{fig:J1356}. The IR flare showed a double peaked structure, with an initial, blue peak, and then a secondary, long-lived, red flare. We obtained optical lightcurves from public, high cadence optical surveys, including ZTF, ASASSN, and ATLAS, and found an optical transient detected in ASASSN data ${\sim}8$ years before the VLASS detection. This transient coincided with the first IR transient peak.

The host galaxy of this transient is shown in the left panel of Figure~\ref{fig:J1356}. Note that this image was taken during the transient, so the bright nucleus is the optical emission associated with the TDE. We obtained an optical spectrum of the host galaxy, which is described in Section~\ref{sec:optspec}. The spectrum is that of a star-forming galaxy. There are no obvious transient spectral features. There is no optical spectrum available at times closer to the optical peak.

\subsection{Optical and IR analysis}
We analyzed the optical flare as described in Section~\ref{sec:optlc}. We fit the combined optical-IR SED at the peak of the IR flare to a blackbody and found a temperature ${\sim}10^4$ K (Figure~\ref{fig:J1356_IR}), consistent with an optically-flaring TDE. The shape of the optical flare is also consistent with TDE observations (see Section~\ref{sec:optlc}).

The secondary IR peak is reminiscent of the ``dust echoes'' occasionally seen from TDEs \citep[][]{tde_dustecho}. Dust echoes are produced by thermally emitting dust in the vicinity of the SMBH, which is heated by UV emission from the TDE. The temperature of the emission is generally ${\sim}1000$ K, i.e., close to but below the sublimation temperature of the dust. To test this hypothesis and constrain the physical properties of the transient, we perform a detailed analysis of this secondary IR flare. We began by fitting blackbody models to each epoch of the secondary flare with $>3\sigma$ detections in both the W1 and W2 bands.

If this flare is a dust echo, the luminosity and timescale of the IR flare is set by the UV lightcurve and the geometry of the dust. Applying simple assumptions, we can set constraints on the dust structure. The secondary IR flare from VT J1356 rose to a peak ${\sim}1200$ days post-optical peak. Assuming the UV flare traces the optical flare and that the dust is in a spherical or torus-like geometry with radius $R_{\rm dust}$, this then constrains $R_{\rm dust}$: the rise time $t_{\rm IR,rise}$ is given by $t_{\rm IR,rise} \sim 2R_{\rm dust}/c$. Then, we have $R_{\rm dust} \sim 0.5$ pc, which is comparable to values found for similar sources \citep[e.g.][]{Somalwar2021}. 

\subsection{Radio and X-ray observations and analysis}

We searched for archival X-ray flares using the ESA high energy lightcurve tool. We obtained X-ray follow-up of this source ${\sim}3000$ days post-peak using the Swift/XRT telescope. We reduced the observations using the online Swift/XRT data analysis tool. No emission was detected ($L_X \lesssim 6\times10^{41}$ erg s$^{-1}$). 

In addition to the VLASS observations of this source, we crossmatched against the Rapid ASKAP Continuum Survey (RACS) catalog \citep{Hale2021PASA...38...58H}. The source was detected on MJD 58600 with a 887.5 MHz flux density of $3.5 \pm 0.9$ mJy. We also obtained a follow-up radio SED, as described in Section~\ref{sec:radioobs}. We fit it to a synchrotron emission model as described in Appendix~\ref{app:synch}. 

We begin by discussing the radio lightcurve. We combined the VLASS, RACS, and follow-up data to construct 887.5 MHz and 3 GHz radio lightcurves, as shown in the top left panel of Figure~\ref{fig:J1356}.

\subsection{The origin of VT J1356}

In this section, we constrain the origin of VT J1356: is it in fact a TDE? We consider three plausible origins to explain these observations: (1) an AGN flare, (2) a supernova (SN), or (3) a TDE. While exotic models may be able to replicate the data, constraining them is beyond the scope of this work. 

We immediately exclude the possibility that we are observing an AGN flare. HG J1356 shows no evidence for AGN activity, either in the form of AGN-like infrared colors nor strong forbidden line emission. If we are observing an AGN flare, it is from an extraordinarily weak AGN, so we consider this scenario unlikely.

Next, we consider a supernova origin. HG J1356 is an actively star forming galaxy, so this possibility is more feasible. Our main evidence against a SN origin comes from the MIR emission. \cite{Yang2019ApJ} analyzed NEOWISE lightcurves \citep[][]{neowise} for 2812 SNe from the Open Catalogue of Supernovae \citep[][]{Guillochon2015} and found that (1) the MIR variability typically occured on a ${\sim}6$ month timescale and (2) SNe MIR lightcurves tend to show a bluer-when-brighter behavior. Our source is variable on many year timescales, and the MIR color becomes redder when the flare is more luminous. As in the AGN flare case, if this event is an SN, it is a very unusual example of such, so we do not prefer this possibility.

We are left with a TDE origin for VT J1356. The optical flare, IR flare, and energetics are consistent with a TDE. 

\section{Optical spectrum fitting methods} \label{app:optspec}

We fit the stellar continua of the spectra using the \texttt{ppxf} code \citep[][]{Cappellari2004, Cappellari2017} with the MILES templates \citep[][]{Vazdekis2010} following the method detailed in Appendix B of \cite{Somalwar2021}. The best fit stellar continua are overlaid on Figure~\ref{fig:opt_specs}. We then subtracted the stellar continuum from each spectrum and fit the individual emission lines to measure the fluxes. We jointly model lines that are closer together than a few times the instrument resolution to ensure reliable fits, such as the [N\,II]+H$\alpha$ complex. We assumed each line could be modelled by a Gaussian. We allowed the amplitude, width, and centroid of each line to float, although when simultaneously modelling multiple lines, we assumed the widths were the same for every line. We included a linear component to account for any residual continuum. For each line or line complex, we ran \texttt{emcee} for $2000$ steps with $200$ walkers and discarded the first $1000$ steps as burn-in.

\section{Synchrotron analysis methodology} \label{app:synch}

We consider an electron in a homogeneous region at redshift $z$ and luminosity distance $d_L$. The rest-frame frequency $\nu$ is related to the observed frequency as $\nu = (1+z)\nu_{\rm obs}$. We will perform most calculations using rest-frame frequency, and transform to observer frequency before comparing to observations. The magnetic field is given by $B$. Consider a single electron with Lorentz factor (LF) $\gamma$. The pitch angle, or the angle between the electron's velocity and the magnetic field, is $\theta$. The rest-frame synchrotron frequency is
\begin{equation}
    \nu_s = \frac{eB \gamma^2}{2\pi mc}.
\end{equation}
The synchrotron power for a single electron at rest-frame frequency $\nu$ is given by
\begin{equation}
    P_s(\nu \mid B, \gamma, \theta) = \frac{\sqrt{3}e^2B\sin\theta}{m_e c^2} F(\nu / \nu_c), \nu_c = \frac{3}{2} \nu_s \sin \theta.
    \label{eq:power_single}
\end{equation}
The function $F(\nu/\nu_c)$ encapsulates the frequency dependence of the spectrum. It is defined as
\begin{equation}
    F(x) = x\int_{x}^{\infty} K_{5/3}(y)dy.
\end{equation}
$K_{5/3}(y)$ is the modified Bessel function of order $5/3$. 

At low and high frequencies, $F(x)$ is well approximated by
\begin{gather}
    F(x) \longrightarrow F_1(x) = \frac{4\pi}{\sqrt{3}\Gamma(1/3)} \bigg( \frac{x}{2} \bigg)^{1/3}, x \rightarrow 0; \\
    F(x) \rightarrow F_2(x) = \sqrt{\frac{\pi}{2}} x^{1/2} e^{-x},  x \rightarrow \infty.
\end{gather}
Here, $\Gamma$ is the Gamma function, rather than bulk Lorentz factor. For all equations hereafter, an expression of the form $\Gamma(x)$ refers to the Gamma function, an expression of the form $\Gamma(s, x)$ refers to the upper incomplete Gamma function, and the character $\Gamma$ with no argument refers to the bulk Lorentz factor. For $x \leq 10^{-5}$ and $x \geq 10^3$, these formulae give relative errors $<0.1\%$. For the regime $10^{-5}<x<10^3$, $F(x)$ is well approximated (relative error $<0.8\%$) as
\begin{gather}
    F(x) \approx F^{(1)}(x) = F_1(x) \frac{\sum_{i=0}^{14} n_{1i} e^{-\alpha_{1i} x}}{\sum_{i=0}^{14} n_{1i}} + F_2(x) \bigg(1 - \frac{\sum_{i=0}^{14} n_{2i} e^{-\alpha_{2i} x}}{\sum_{i=0}^{14} n_{1i}} \bigg)
    \label{eq:F_approx1}
\end{gather}
Note that the factor $\sum_{i=0}^{14} n_{1i}$ is the denominator of the final term is {\it not} a typo. Because we use the sum over $n_{1i}$ in the denominator, $F^{(1)}(x)$ does not approach $F(x)$ exactly as $x$ goes to infinity. However, as we will show, the difference does not affect our results: for all upcoming calculations, this error will prove insignificant.

The single electron synchrotron spectrum, Equation~\ref{eq:power_single}, is only valid when the emitting energy during a single orbit is smaller than the energy of the particle, in which regime quantum effects are negligible. The condition corresponds to 
\begin{equation}
    B < \frac{e/\sigma_{\rm T}}{\gamma^2 \sin^2\theta} \sim \frac{7.22 \times 10^{14}}{\gamma^2 \sin^2 \theta}\,{\rm G},
\end{equation}
where $\sigma_{\rm T}$ is the Thomson cross section. 

To compute the synchrotron spectrum for a population of electrons, we must assume an electron energy distribution. We adopt the standard assumption that the electron lorentz factors are drawn from a cut-off power-law with index $-p$
\begin{gather}
    N(\gamma)d\gamma = N_0 \gamma^{-p} d\gamma\,\,(\gamma_{\rm min} < \gamma < \gamma_{\rm max}) \\
    = N_{\rm tot} \frac{1-p}{\gamma_{\rm max}^{1-p} - \gamma_{\rm min}^{1-p}} \gamma^{-p} d\gamma\,\,(\gamma_{\rm min} < \gamma < \gamma_{\rm max}) \\
    = E_{\rm e} \frac{2-p}{\gamma_{\rm max}^{2-p} - \gamma_{\rm min}^{2-p}} \gamma^{-p} d\gamma\,\,(\gamma_{\rm min} < \gamma < \gamma_{\rm max}).
\end{gather}
$\gamma_{\rm min (max)}$ is the minimum (maximum) electron Lorentz factor. A common assumption is $\gamma_{\rm max} = \infty$. Throughout the rest of this work, we assume $\gamma_{\rm max} = 10^9$, which is roughly equivalent to $\gamma_{\rm max} = \infty$, although all derivations are generalized to arbitrary $\gamma_{\rm max}$. $N_0$ is the normalization of the electron distribution, and is related to both the total number of electrons $N_{\rm tot} = \int_{\gamma_{\rm min}}^{\gamma_{\rm max}} N(\gamma) d\gamma$ and the total energy stored in the electrons $E_{e} = \int_{\gamma_{\rm min}}^{\gamma_{\rm max}} \gamma m_e c^2 N(\gamma) d\gamma$.

Using this energy distribution, the synchrotron emissivity (cgs units erg s$^{-1}$ cm$^{-3}$ sterad$^{-1}$) is given by
\begin{gather}
    \epsilon_s(\nu) = \frac{1}{4\pi} \int_{\gamma_{\rm min}}^{\gamma_{\rm max}} d\gamma N(\gamma) P(\nu \mid B, \gamma, \theta) \\
    = \frac{\sqrt{3}}{8\pi} \frac{e^3 N_0 B \sin \theta}{m_e c^2} \bigg(\frac{\nu}{\nu_0}\bigg)^{\frac{1-p}{2}} \int_{x_{\rm max}}^{x_{\rm min}} x^{(3-p)/2} F(x) dx.
    \label{eq:eps}
\end{gather}
Here, we have defined the variable $\nu_0 = \frac{3eB\sin \theta \Gamma}{4\pi m_ec (1+z)}$, so that $\nu_c = \gamma^2 \nu_0$. We have also defined $x_{\rm min} = \nu/(\nu_0 \gamma_{\rm min}^2)$ and $x_{\rm max} = \nu/(\nu_0 \gamma_{\rm max}^2)$; note that, despite the subscripts, $x_{\rm min} > x_{\rm max}$. In the limit $x_{\rm min} \rightarrow \infty$ and $x_{\rm max} \rightarrow 0$, we find the expected frequency dependence $\epsilon_s(\nu) \propto (\nu/\nu_0)^{(1-p)/2}$.

The approximation for $F(x)$ described above allows this integral to be analytically evaluated. We find
\begin{gather}
    \epsilon^{(2)}_s(\nu) = \frac{\sqrt{3}}{8\pi} \frac{e^3 N_0 B \sin \theta}{m_e c^2} \bigg(\frac{\nu}{\nu_0}\bigg)^{\frac{1-p}{2}} \Bigg\{ F_1 \frac{\sum_{i=0}^{14} n_{1i} \alpha_{1i}^{-z_1} \big(\Gamma(z_1, \alpha_{1i}x_{\rm min}) - \Gamma(z_1, \alpha_{1i}x_{\rm max}) \big)}{\sum_{i=0}^{14} n_{1i}}  +  \nonumber  \\ 
    F_2 \bigg[ \big(\Gamma(z_2, x_{\rm min}) - \Gamma(z_2, x_{\rm max}) \big) - \frac{\sum_{i=0}^{14} n_{2i} (\alpha_{2i}+1)^{-z_2} \big(\Gamma(z_2, (\alpha_{2i}+1)x_{\rm min}) - \Gamma(z_2, (\alpha_{2i}+1)x_{\rm max}) \big)}{\sum_{i=0}^{14} n_{1i}} \bigg] \Bigg\}; \label{eq:eps_approx1} \\ \nonumber
    z_1 = \frac{3-p}{2} + \frac{1}{3} + 1, z_2 = \frac{3-p}{2} + \frac{1}{2} + 1.
\end{gather}
The relative error of this expression is $<0.1\%$ across the full parameter range.

With these expressions for the emissivity, we can readily calculate the synchrotron spectrum for any {\it optically thin} emission region. In many cases, however, we must consider absorption processes. First, we consider synchrotron self-absorption. The synchrotron self-absorption coefficient is given by
\begin{gather}
    \alpha_\nu = \frac{1}{8 \pi m_e \nu^2} \int_{\gamma_{\rm min}}^{\gamma_{\rm max}} d\gamma P_s(\nu) \gamma^2 \frac{\partial}{\partial \gamma } \bigg[ \frac{N(\gamma)}{\gamma^2} \bigg]
    = \frac{1}{8 \pi m_e \nu^2} \int_{\gamma_{\rm min}}^{\gamma_{\rm max}} d\gamma \frac{N(\gamma)}{\gamma^2} \frac{\partial}{\partial \gamma} \bigg[ \gamma^2 P_s(\nu, \theta) \bigg],
\end{gather}
where the second expression uses the fact that $N(\gamma_{\rm min}) = N(\gamma_{\rm max}) = 0$, and is particularly relevant in cases where the derivative of $N(\gamma)$ is not well-defined. Using our power-law expression for $N(\gamma)$, we find
\begin{gather}
    \alpha_\nu = \frac{(p+2) \sqrt{3} e^3 N_0 B \sin \theta}{16 \pi \nu^2 m_e^2 c^2} \bigg(\frac{\nu}{\nu_0}\bigg)^{-\frac{p}{2}} \int_{x_{\rm min}}^{x_{\rm max}} dx\,x^{p/2-1} F(x)
\end{gather}
In the limit $x_{\rm min}\rightarrow\infty$ and $x_{\rm max}\rightarrow0$, we find $\alpha_\nu \propto \nu^{-(p+4)/2}$, as expected.

As before, we can analytically evaluate this expression using our approximations. Evaluating for $F^{(1)}(x)$, we find
\begin{gather}
    \alpha^{(1)}_\nu = \frac{(p+2) \sqrt{3} e^3 N_0 B \sin \theta}{16 \pi \nu^2 m_e^2 c^2} \bigg(\frac{\nu}{\nu_0}\bigg)^{-\frac{p}{2}}
    \Bigg\{ F_1 \frac{\sum_{i=0}^{14} n_{1i} \alpha_{1i}^{-z_1} \big(\Gamma(z_1, \alpha_{1i}x_{\rm min}) - \Gamma(z_1, \alpha_{1i}x_{\rm max}) \big)}{\sum_{i=0}^{14} n_{1i}}  +  \nonumber  \\ 
    F_2 \bigg[ \big(\Gamma(z_2, x_{\rm min}) - \Gamma(z_2, x_{\rm max}) \big) - \frac{\sum_{i=0}^{14} n_{2i} (\alpha_{2i}+1)^{-z_2} \big(\Gamma(z_2, (\alpha_{2i}+1)x_{\rm min}) - \Gamma(z_2, (\alpha_{2i}+1)x_{\rm max}) \big)}{\sum_{i=0}^{14} n_{1i}} \bigg] \Bigg\}; \label{eq:anu_approx1}\\ \nonumber
    z_3 = \frac{2-p}{2} + \frac{1}{3} + 1, z_4 = \frac{2-p}{2} + \frac{1}{2} + 1.
\end{gather}
The relative error in this expression is $<0.1\%$ across the full parameter space.

We are now in a position to solve the radiative transfer equation to calculate the observed spectrum for an arbitrary source. For reference, we restate the radiative transfer equation here:
\begin{equation}
    \frac{dI_\nu}{ds} = \alpha_\nu I_\nu + \epsilon_s.
\end{equation}
Here, $I_\nu$ is the specific intensity [erg cm$^{-2}$ s$^{-1}$  Hz$^{-1}$ sterad$^{-1}$] and $s$ is the path length. $\alpha_\nu$ and $\epsilon_s$ are still, respectively, the absorption coefficient [cm$^{-1}$] and emissivity [erg cm$^{-3}$ s$^{-1}$ sterad$^{-1}$]. For a sphere with radius $R$ ($A=\pi R^2$, $s=2R$), the flux density is given by
\begin{equation}
    F_{\nu, {\rm sphere}} = \frac{f_A A}{d_L^2} \frac{\epsilon_s(\nu)}{\alpha_\nu(\nu)} \Bigg[ 1 - \frac{2}{\tau_\nu^2} \bigg( 1 - (1 + \tau_\nu) e^{-\tau_\nu} \bigg) \Bigg];\,\,\,\tau_{\rm sa}(\nu) = s \alpha_\nu(\nu).
\end{equation}

We adopt this approximation for the synchrotron SED for all the analysis in this work.

{\addtolength{\tabcolsep}{-2pt}\startlongtable   \begin{deluxetable*}{c | ccccccccc}
\centerwidetable
\tablewidth{10pt}
\tablecaption{TDEs with published radio SEDs  \label{tab:radio_sample}}
\tablehead{ \colhead{Name} & \colhead{$t_{\rm obs}-t_{\rm opt. peak}$} & \colhead{$\log R/{\rm cm}$} & \colhead{$\log B/{\rm G}$} & \colhead{$\log N_0 / {\rm cm^{-3}}$} & \colhead{$p$} & \colhead{$\beta$} & \colhead{$\log E$} & \colhead{$\log M_{\rm ej}/M_\odot$} & \colhead{SED Source} }
\startdata
 VT J0813 & 17  & $18.2_{-0.9}^{+0.8}$ & $-2.0_{-0.7}^{+0.8}$ & $-2_{- 1}^{+ 1}$ & $1.69_{-0.05}^{+0.06}$ & $-0.03_{-0.16}^{+0.03}$ & $50.4_{-1.0}^{+1.0}$ & $-3.5_{-1.0}^{+1.0}$ \cr 
 & 45  & $17.2_{-0.7}^{+1.2}$ & $-1.4_{-1.0}^{+0.6}$ & $ 1_{- 2}^{+ 2}$ & $2.08_{-0.08}^{+0.13}$ & $-0.4_{-0.5}^{+0.3}$ & $48.4_{-1.0}^{+1.6}$ & $-5.5_{-1.0}^{+1.6}$ \cr 
 & 53  & $17.7_{-0.8}^{+0.9}$ & $-1.8_{-0.8}^{+0.6}$ & $ 0_{- 2}^{+ 1}$ & $2.06_{-0.10}^{+0.11}$ & $-0.2_{-0.4}^{+0.1}$ & $49_{- 1}^{+ 1}$ & $-5_{- 1}^{+ 1}$ \cr 
 & 83  & $18.1_{-0.6}^{+0.7}$ & $-2.2_{-0.6}^{+0.5}$ & $-0_{- 1}^{+ 1}$ & $2.20_{-0.03}^{+0.03}$ & $-0.09_{-0.22}^{+0.07}$ & $49.6_{-0.8}^{+0.9}$ & $-4.3_{-0.8}^{+0.9}$ \cr 
 & 134  & $17.05_{-0.01}^{+0.01}$ & $-1.17_{-0.02}^{+0.02}$ & $2.35_{-0.06}^{+0.06}$ & $2.55_{-0.03}^{+0.03}$ & $-0.68_{-0.01}^{+0.01}$ & $48.49_{-0.02}^{+0.03}$ & $-5.46_{-0.02}^{+0.03}$ \cr 
 & 205  & $17.169_{-0.006}^{+0.007}$ & $-1.15_{-0.01}^{+0.01}$ & $2.56_{-0.04}^{+0.03}$ & $2.81_{-0.03}^{+0.02}$ & $-0.712_{-0.005}^{+0.006}$ & $48.88_{-0.02}^{+0.02}$ & $-5.07_{-0.02}^{+0.02}$ \cr 
 & 247  & $17.2_{-0.2}^{+1.4}$ & $-1.3_{-1.1}^{+0.3}$ & $2.2_{-2.3}^{+0.6}$ & $2.57_{-0.09}^{+0.16}$ & $-0.7_{-0.2}^{+0.7}$ & $48.8_{-0.2}^{+1.9}$ & $-5.1_{-0.2}^{+1.9}$ \cr 
 & 302  & $17.00_{-0.05}^{+0.11}$ & $-1.10_{-0.11}^{+0.06}$ & $2.3_{-0.3}^{+0.2}$ & $2.38_{-0.10}^{+0.08}$ & $-0.98_{-0.04}^{+0.10}$ & $48.48_{-0.04}^{+0.10}$ & $-5.47_{-0.04}^{+0.10}$ \cr 
 & 410  & $17.25_{-0.01}^{+0.02}$ & $-0.86_{-0.02}^{+0.02}$ & $3.42_{-0.05}^{+0.05}$ & $3.51_{-0.03}^{+0.03}$ & $-0.86_{-0.01}^{+0.01}$ & $49.73_{-0.02}^{+0.02}$ & $-4.23_{-0.02}^{+0.02}$ \cr 
 & 618  & $17.341_{-0.005}^{+0.005}$ & $-1.237_{-0.008}^{+0.008}$ & $2.29_{-0.02}^{+0.02}$ & $2.65_{-0.01}^{+0.01}$ & $-0.933_{-0.005}^{+0.005}$ & $49.221_{-0.008}^{+0.008}$ & $-4.730_{-0.008}^{+0.008}$ \cr 
 & 701  & $17.43_{-0.01}^{+0.01}$ & $-1.11_{-0.02}^{+0.01}$ & $2.80_{-0.04}^{+0.03}$ & $3.19_{-0.02}^{+0.02}$ & $-0.90_{-0.01}^{+0.01}$ & $49.74_{-0.01}^{+0.01}$ & $-4.21_{-0.01}^{+0.01}$ \cr 
 & 800  & $17.23_{-0.02}^{+0.02}$ & $-1.15_{-0.04}^{+0.04}$ & $2.6_{-0.1}^{+0.1}$ & $2.85_{-0.06}^{+0.07}$ & $-1.13_{-0.02}^{+0.02}$ & $49.08_{-0.04}^{+0.05}$ & $-4.87_{-0.04}^{+0.05}$ \cr\hline
 VT J1008 & 507  & $17.06_{-0.01}^{+0.01}$ & $-0.73_{-0.04}^{+0.05}$ & $3.5_{-0.1}^{+0.2}$ & $3.0_{-0.1}^{+0.1}$ & $-1.37_{-0.01}^{+0.01}$ & $49.4_{-0.1}^{+0.1}$ & $-4.5_{-0.1}^{+0.1}$ \cr\hline
 VT J1356 & 2581  & $17.40_{-0.04}^{+0.04}$ & $-1.12_{-0.03}^{+0.03}$ & $0.2_{-0.2}^{+0.2}$ & $1.80_{-0.04}^{+0.04}$ & $-1.73_{-0.04}^{+0.04}$ & $49.7_{-0.1}^{+0.1}$ & $-4.3_{-0.1}^{+0.1}$ \cr\hline
 VT J1752 & 756  & $17.814_{-0.004}^{+0.003}$ & $0.629_{-0.013}^{+0.007}$ & $-2.75_{-0.02}^{+0.03}$ & $1.004_{-0.003}^{+0.007}$ & $-0.819_{-0.003}^{+0.003}$ & $54.38_{-0.04}^{+0.02}$ & $0.43_{-0.04}^{+0.02}$ \cr 
 & 1115  & $17.99_{-0.02}^{+0.02}$ & $-0.87_{-0.05}^{+0.05}$ & $-0.5_{-0.2}^{+0.2}$ & $1.64_{-0.04}^{+0.04}$ & $-0.75_{-0.02}^{+0.02}$ & $51.9_{-0.1}^{+0.1}$ & $-2.0_{-0.1}^{+0.1}$ \cr\hline
 VT J2012 & 469  & $17.225_{-0.006}^{+0.006}$ & $-0.83_{-0.02}^{+0.02}$ & $3.27_{-0.08}^{+0.08}$ & $2.95_{-0.06}^{+0.07}$ & $-1.178_{-0.006}^{+0.006}$ & $49.69_{-0.05}^{+0.05}$ & $-4.26_{-0.05}^{+0.05}$ \cr\hline
 VT J2030 & 675  & $17.44_{-0.01}^{+0.01}$ & $-0.94_{-0.02}^{+0.02}$ & $3.09_{-0.06}^{+0.06}$ & $3.03_{-0.04}^{+0.04}$ & $-1.122_{-0.010}^{+0.011}$ & $50.11_{-0.03}^{+0.03}$ & $-3.84_{-0.03}^{+0.03}$ \cr 
 & 934  & $17.622_{-0.009}^{+0.009}$ & $-1.02_{-0.02}^{+0.02}$ & $3.04_{-0.04}^{+0.04}$ & $3.31_{-0.03}^{+0.03}$ & $-1.021_{-0.008}^{+0.008}$ & $50.51_{-0.03}^{+0.03}$ & $-3.44_{-0.03}^{+0.03}$ \cr 
 & 1051  & $17.51_{-0.01}^{+0.01}$ & $-1.05_{-0.03}^{+0.03}$ & $2.89_{-0.07}^{+0.08}$ & $3.07_{-0.06}^{+0.06}$ & $-1.15_{-0.01}^{+0.01}$ & $50.11_{-0.04}^{+0.04}$ & $-3.84_{-0.04}^{+0.04}$ \cr\hline
 \enddata
 \tablecomments{The physical parameters of the radio emission regions for our TDE candidates.}
\centering
\end{deluxetable*}}

\end{appendix}

\end{document}